# Technical Report: A Methodology for Studying 802.11p VANET Broadcasting Performance with Practical Vehicle Distribution

Harry J. F. Qiu*, Ivan Wang-Hei Ho†, Chi K. Tse†, Fellow, IEEE, Yu Xie†
*Intelligent Assistive Technology and Systems Lab, University of Toronto
†Department of Electronic and Information Engineering, The Hong Kong Polytechnic University
jf.qiu@mail.utoronto.ca, {ivanwh.ho, michael.tse}@polyu.edu.hk, yu.xie@connect.polyu.hk

**Abstract –** In a Vehicular Ad-hoc Network (VANET), the performance of the communication protocol is influenced heavily by the vehicular density dynamics. However, most of the previous works on VANET performance modeling paid little attention to vehicle distribution, or simply assumed homogeneous car distribution. It is obvious that vehicles are distributed non-homogeneously along a road segment due to traffic signals and speed limits at different portions of the road, as well as vehicle interactions that are significant on busy streets. In light of the inadequacy, we present in this paper an original methodology to study the broadcasting performance of 802.11p VANETs with practical vehicle distribution in urban environments. Firstly, we adopt the empirically verified stochastic traffic models, which incorporates the effect of urban settings (such as traffic lights and vehicle interactions) on car distribution and generates practical vehicular density profiles. Corresponding 802.11p protocol and performance models are then developed. When coupled with the traffic models, they can predict broadcasting efficiency, delay, as well as throughput performance of 802.11p VANETs based on the knowledge of car density at each location on the road. Extensive simulation is conducted to verify the accuracy of the developed mathematical models with the consideration of vehicle interaction. In general, our results demonstrate the applicability of the proposed methodology on modeling protocol performance in practical signalized road networks, and shed insights into the design and development of future communication protocols and networking functions for VANETs.

## 1. INTRODUCTION

With the inception of Intelligent Transportation System (ITS), the demand of information exchange between vehicles and infrastructure (V2I) or among vehicles (V2V) is on the rise. By extending information delivery to automobiles, ITS has the potential to facilitate efficient traffic routing, provide infotainment service, and most importantly enhance road safety [1, 2].

Government, academic and industry have been paving the way to materialize ITS in the past two decades. In 1999, the Federal Communications Commission (FCC) allocated a 75 MHz Dedicated Short Range Communication (DSRC) band centered at 5.9 GHz for vehicular communications [3]. In 2010, the IEEE 802.11p amendment [4] was approved for the underlying radio technology for DSRC, which extended the existing 802.11 standard to cater the requirements resulting from the highly-mobile nature of VANET.

802.11p attracted tremendous amount of research interest, a majority of which focused on analytical modeling and performance evaluation of the protocol. As cars become the carriers of 802.11p transceivers, there is a key issue on how to incorporate vehicular traffic into the analytical model and performance evaluation. The existing methodologies for studying VANET performance can primarily be divided into two groups. One group provides analytical models of 802.11p based on simplified assumptions of the vehicular traffic, such as homogeneous car distribution. References [5, 6, 7, 8, 9, 10, 11] fall into this category. By omitting important factors of urban road settings, such as traffic signals and car-to-car interaction, these models failed to capture the spatial and temporal variations of network performance along a road in the real-world setting. The other group of research [12, 13, 14, 15, 16] relies entirely on simulation to generate traffic profile as well as network communication traffic for performance evaluation. Although it provides reliable performance indicators, this approach is computationally expensive and hence not suitable for quick performance prediction. For example, simulating a traffic network with 3600 cars for 6 minutes took as long as 30 hours [13]. Both groups of research will be explained in further details in the next section.

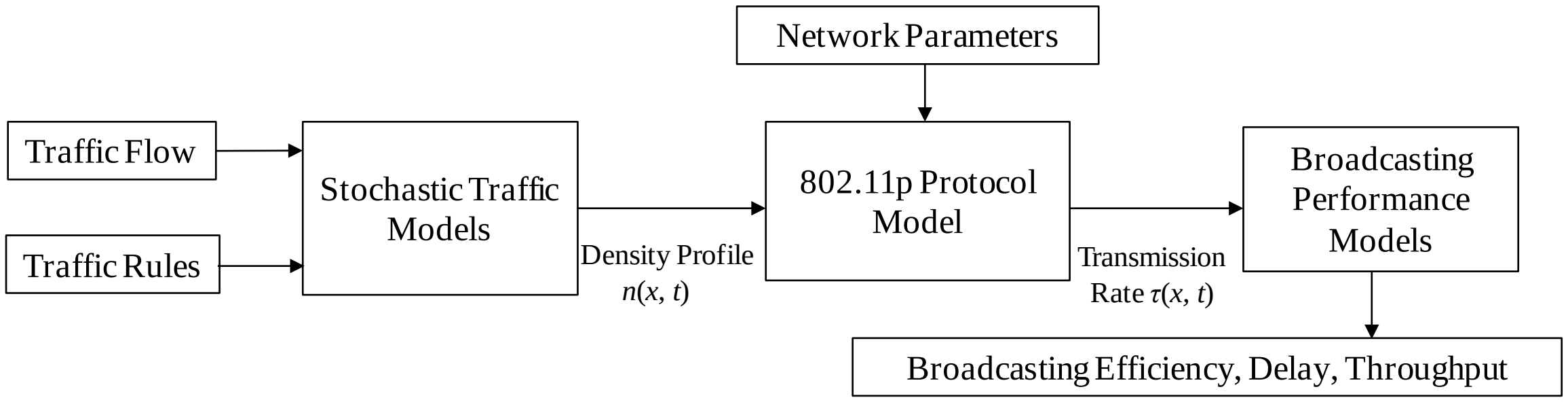

Figure 1.1. An overview of the methodology for 802.11p VANET broadcasting performance modeling.

Our work uniquely contributes to the state of the art by providing a novel methodology to study 802.11p VANET performance with practical vehicle distribution. Figure 1.1 provides an overview of our methodology. Firstly, we modified the stochastic traffic models in [17] for the following protocol performance analysis and build on top of it the 802.11p protocol and performance models to provide analysis on broadcasting performance (in terms of packet delivery success rate, delay and throughput). Specifically, we considered beacon broadcasting in the control channel (CCH) to demonstrate this analytical framework. Extensive simulations verified that our results successfully capture the spatial and temporal variations of 802.11p broadcasting performance along a signalized urban road segment. By utilizing the proposed framework, researchers and network designers can readily compute analytical performance results regarding a signalized urban vehicular network, whose car distribution is generated based on either synthetic or empirical vehicular velocity profiles. In addition, our models can be exploited to support network condition prediction and preemptive network configuration (e.g., packet transmission rate and transmission power control) to optimize VANET performance at different road locations dynamically.

The current models extend our previous works [17] and [18]. Compared with [17], which primarily introduces the stochastic traffic models and demonstrates how network connectivity could be predicted from the traffic dynamics, this paper takes an important step forward to consider the effect of packet contention and collision in the context of the 802.11p communication protocol. Furthermore, the preliminary work [18] lacks the consideration of most of the important performance parameters (e.g., packet delay and network throughput) and overlooks some of the practical broadcasting characteristic in a VANET (e.g., the nature of the unsaturated safety packets), this paper provides a more comprehensive illustration of the overall methodology from vehicular traffic modeling to protocol performance estimation.

The rest of the paper is organized as follows. Section 2 provides a comprehensive background and related work review on vehicular traffic modeling and 802.11p protocol models; Section 3 describes the proposed VANET modeling methodology; Section 4 presents the simulation results for verifying our analytical models; Section 5 initiates discussion on the results and provide direction for future research; finally, Section 6 concludes the paper.

## 2. Background and Related Work

To enable a standardized access to the DSRC band, the DSRC protocol stack [3] was devised, in which IEEE 802.11p described the PHY and MAC layers, while the IEEE 1609 (WAVE) standard was responsible for higher-layer issues, such as multi-channel operation, the WAVE short message protocol (WSMP), authentication and encryption, etc. The PHY layer of 802.11p mostly adopts that of 802.11a (e.g., using the OFDM technology to cope with the highly dynamic channel condition), except that a default channel width option of 10 MHz is introduced in addition to the 20 MHz option in 802.11a. It means that the symbol duration is doubled in 802.11p, which helps the system to cope with multipath fading in the vehicular environment.

Table 2.1. Contention parameters for different access categories in 802.11p.

| AC | Data Class | CWmin | CWmax | AIFS |
|---|---|---|---|---|
| 3 | Safety Related | 3 | 7 | 2 |
| 2 | Voice | 3 | 7 | 3 |
| 1 | Best Effort | 7 | 15 | 6 |
| 0 | Background Traffic | 15 | 1023 | 9 |

The MAC layer of 802.11p inherits Distributed Coordination Function (DCF) from 802.11, which employs CSMA/CA with the exponential backoff mechanism. Bianchi [5] pioneered the analytical modelling of DCF with a 2D Markov Chain, and evaluated the node transmission rate and throughput under saturated condition (i.e., a node always has a packet ready for transmission). To further improve the quality of service (QoS), 802.11p adopts the Enhanced Distributed Channel Access (EDCA) from 802.11e [19], which utilizes four access classes (AC) to schedule arrived packets. Packets from different ACs within a node contend internally, and only the winner will participate in external contention. The contention parameters (e.g., the contention window size and arbitrary inter-frame space (AIFS)) shown in Table 2.1 are configured so that highly important messages (such as safety broadcast) fall in AC3 with the lowest AIFS and contention window size are most likely to win the internal contention. Eichler *et al.* [7] provided a simple model based on combinatory theory to describe the delay and throughput of each AC independently; their result verified the effectiveness of EDCA. Moreover, Han *et al*. [6] proposed a mathematically tractable model to describe EDCA under saturated condition based on a 2D Markov chain and contention zone segmentation.

Among all types of messages flowing through a VANET, broadcasting messages transmitted over the control channel (CCH) are crucial for disseminating safety information within a local region, and hence the performance modeling of safety broadcasting attracts a lot of research interest. However, most of the analytical work on 802.11p broadcasting performance evaluation did not provide a realistic account of the dynamics of vehicular traffic. Instead, they either considered a collection of static cars or a segment of highway with homogenous car distribution for the sake of analytical simplicity. Ma *et al.* [8] presented a 1D Markov Chain to describe the saturated broadcast throughput, delay, and packet reception ratio, taking channel freezing into account. As the safety messages arrives infrequently, [9] extended [8] by modelling the unsaturated message queue by a M/G/1 queue, from which the same set of performance indicators were obtained using a probability generating function (PGF). In light of the fact that periodic beacons constitute the majority of the broadcasting traffic, and their arrivals are deterministic, Vinel *et al.* [10] devised a 3D Markov chain model to analyze periodic beacon broadcast performance. In their model, every node was preloaded with a beacon message and the number of active nodes decreased when more nodes finished their transmissions. The resulting Markov chain consists of a large number of states, which was a major drawback of the model. Wang *et al.* [11] adopted a 1D Markov chain similar to that of [8], but they focused on the trade-off between broadcast throughput and reliability, and the Pareto frontiers were identified for the throughput-reliability function, from which contention window trade-off zones were found for different number of nodes in the system.

When studying the performance of 802.11p, the impact of urban traffic on network performance should not be overlooked. Another stream of research resorted to sophisticated traffic simulation for generating realistic car mobility traces, on which network evaluations could be carried out either through simulation or model analysis. Despite their accuracy, such approach faces two shortcomings: i) the simulation running time is too long to support prediction and protocol adjustment in real time; and ii) the simulation result for a particular scenario might not be able to be generalized and applied to others. Jafari *et al.* [12] investigated how relative car positions affected broadcast packet loss rate, end-to-end delay and throughput. They constructed a simple highway scenario in VanetMobiSim [20], where ten cars moved in one direction at three different speed levels; meanwhile, NS-2 [21] was adopted to simulate the message broadcasting process. Grafling *et al.* [14] used VanetMobiSim to simulate traffic flows in three scenarios: open area, free highway and Washington street map. They simulated the 802.11p broadcast in the Qualnet simulator [22], paying specific attention to the multi-channel operation. Their results verified that the control channel (CCH) had better access privilege than other channels, and hence resulted in better delay and throughput performances. Noori *et al.* [13] ran a traffic simulation in SUMO [23] based on an empirical data set collected from the Cologne city, and the beaconing behavior of the network was simulated by OMNeT++ [24]. They verified that the beacon delivery success rate decreased as the number of nodes increased, while the presence of road intersection improved the success rate.

Different from the above, our work provides a framework to analytically evaluate both the 802.11p network performance and the underlying traffic distribution. As such, our analytical models are capable to capture both the spatial and temporal dynamics of various network performance indexes along an urban signalized road segment. To the author's best knowledge, there are two pieces of work [25, 26] that share some similarities with ours.

Ma *et al.* [25] extended their previous work [9] by adding a traffic model component of a highway with homogeneous car distribution. By considering channel freezing, channel fading, hidden nodes and link breaking, they found that hidden nodes notably degraded the broadcast performance while the impact of link breaking was insignificant. At the same time, broadcast successful reception rate and delay deteriorated with higher traffic density. Compared to [25], our work provides a more realistic network evaluation over heterogeneous traffic distribution in a signalized road system. In addition, we developed our own unsaturated model to characterize the periodic beacon broadcasting, which could not be described by the M/G/1 queue model in [25].

Bastani *et al.* [26, 27] developed a heterogeneous traffic model that considered traffic light at a road junction, based on the simulation result from the Paramics traffic simulator [28]. Furthermore, a network model similar to that of [9] was employed to model the VANET broadcast behavior. In the traffic model, the region before the traffic light was divided into three regions: the free run, decelerating, and stop regions; then, a logistic function whose parameters were obtained according to the motion dynamics in each region was devised. Although [26] also captured the spatial and temporal variations of cars distribution, it differs from our model in three aspects: i) it is a kind of simulation trace-based model, hence the generalization of the piece-wise logistic density function to other traffic scenarios is limited (e.g., from a road junction to more complicated road networks); on the other hand, our model is developed via a purely analytical approach, its adaptation to other traffic scenarios (e.g., curved road, 3D road networks, etc.) can be easily achieved by providing the corresponding velocity profile in terms of space and time; ii) the traffic model proposed in [27] is deterministic in nature based on the simulation result, in contrast, our model characterizes both the average and distributional result for the vehicular traffic via the fluid model and stochastic model respectively; and iii) Our model provides higher spatial and temporal details by considering the car-to-car interaction rigorously via the density-dependent velocity profile, such details were simplified in Bastania's model via the piece-wise logistic density function.

## 3. Methodology for Performance Modeling of 802.11p VANETs

In this section, we propose an original methodology for modeling the performance of 802.11p VANETs under practical vehicle distribution, which consists of the stochastic traffic models that capture the time and space dynamics of the vehicular traffic, and the corresponding 802.11p protocol and performance models for estimating various network performance based on the vehicular density profile.

### 3.1 Stochastic Traffic Models

The stochastic traffic model combines a macroscopic flow model with a stochastic model, which captures the randomness in the system. We also consider vehicle interactions through a density-dependent velocity profile, enabling network performance prediction of non-homogeneously distributed traffic. A road segment in Figure 3.1 with semi-infinite length is considered, and a one-dimensional coordinate system is established with the origin located at the left most point and the positive direction points to the right. Cars arrive at location 0 from the left according to an arrival process $\{A(t) | -\infty < t < +\infty\}$, with non-negative and integrable arrival rate $\alpha(t)$ $(t > 0)$. In the road configuration, the traffic flows in the arrow direction, and each junction has a traffic light that regulates the traffic, which is the only place where cars can exit or enter the road segment. More complex scenarios, such as multi-lane, bi-directional traffic can be constructed by superposing multiple building blocks.

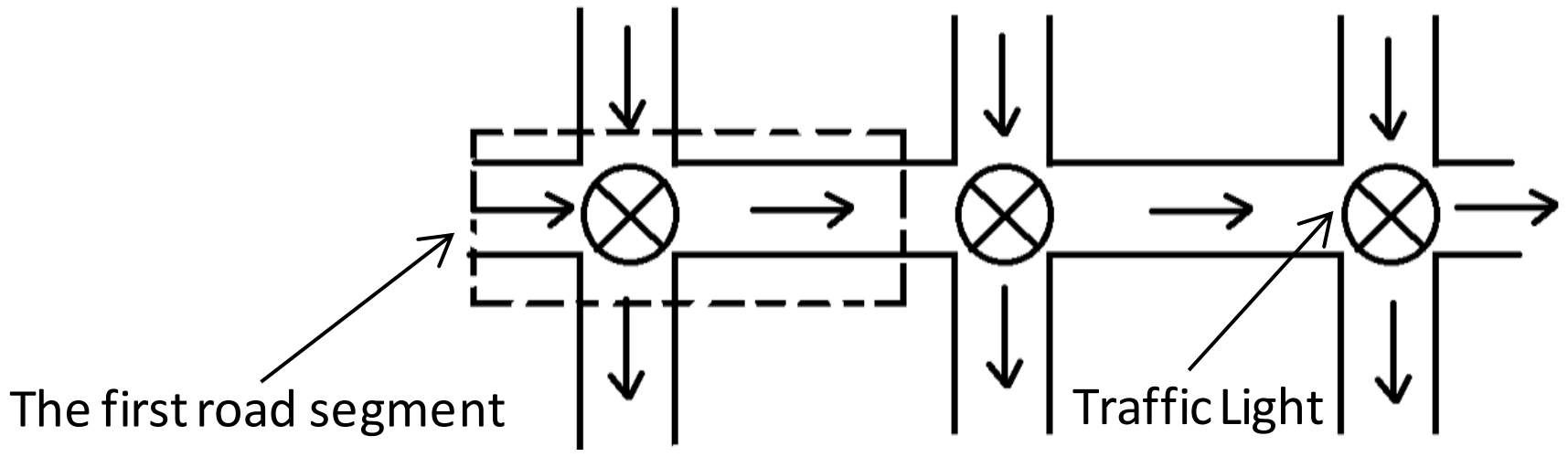

Figure 3.1. The road scenario considered.

Table 3.1. Summary of notations and symbols for the stochastic traffic models.

| | |
|---|---|
| $\alpha(t)$ | Car arrival rate as a function of time $t$ |
| $C^+(x,t)$ | The number of cars entering the road segment $(0,x]$ during time interval $(-\infty,t)$ |
| $C^-(x,t)$ | The number of cars leaving the road segment $(0,x]$ during time interval $(-\infty,t)$ |
| $Q(x,t)$ | The number of cars passing the point $x$ within $(-\infty,t)$ |
| $N(x,t)$ | The number of cars remaining in the road segment $(0,x]$ at time $t$ |
| $c^+(x,t)$ | The rate of change of car density with respect to $C^+(x,t)$ |
| $c^-(x,t)$ | The rate of change of car density with respect to $C^-(x,t)$ |
| $q(x,t)$ | the rate at which cars pass through location $x$ at time $t$ |
| $n(x,t)$ | The car density at location $x$ at time $t$ |
| $v(x,t)$ | The velocity of cars at location $x$ at time $t$ |
| $v_f$ | The mean free speed of cars |
| $k_j$ | The congestion density |
| $\sigma(x,t)$ | the time instant at which the car arrived will remain in the road segment $(0,x]$ at time $t$ |

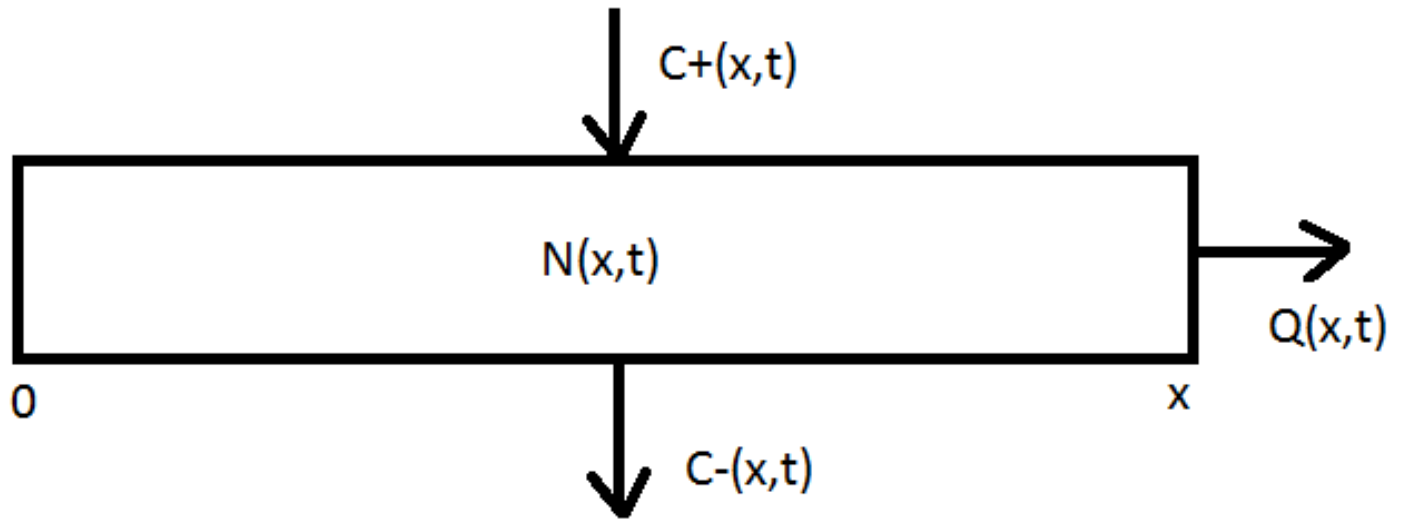

Figure 3.2. Fluid model of traffic flow.

Table 3.2. Assumptions for the stochastic traffic models.

1. Cars entering the road segment from the origin according to a non-homogeneous Poisson arrival process with rate $\alpha(t), t > 0$
2. There is no car leaving the road segment except at location $x$
3. The velocity of a car $v(x,t)$ is independent of the local car density $n(x,t)$

By singling out the first road segment in Figure 3.1, the traffic within can be viewed as incompressible fluid (Figure 3.2). Our model aims at inferring the spatial and temporal distribution of cars along a road, *i.e.*, $N(x,t)$, based on measurable inputs such as the inflow $(C^+)$ to and the outflows $(C^-, Q)$ from the road segment. The notations and assumptions for the traffic models are summarized in Table 3.1 and Table 3.2 respectively.

According to the fluid conservation principle, and applying the fundamental relationship among flow rate, flow speed and fluid density, we can derive the following differential equations

$$c^{+}(x,t) - c^{-}(x,t) - \frac{\partial q(x,t)}{\partial x} = \frac{\partial n(x,t)}{\partial t} \quad (1)$$

$$c^{+}(x,t) - c^{-}(x,t) - \frac{\partial v(x,t)}{\partial x} n(x,t) = \frac{dn(x,t)}{dt} \quad , \quad (2)$$

where $c^{+}(x,t) = \frac{\partial^2 C^{+}(x,t)}{\partial x \partial t}$ denotes the rate of change of car entering density; $c^{-}(x,t) = \frac{\partial^2 C^{-}(x,t)}{\partial x \partial t}$ denotes the rate of car exiting density; $q(x,t) = \partial Q(x,t)/\partial t$ is the rate at which cars pass through location *x*. Hence, $\partial q(x,t)/\partial x$ is the spatial variation of the flow rate, and $\partial n(x,t)/\partial t$ is the rate of change of car density in the road segment. By assuming that the velocity $v(x,t)$ is independent of the car density $n(x,t)$, the partial differential equation (PDE) (1) can be converted into an ordinary differential equation (ODE) (2), which is easier to solve.

Therefore, if the vehicle arrival and departure patterns ($c^{+}$ and $c^{-}$) and the velocity distribution $v(x,t)$ are known for a road segment, then the vehicular density can be readily obtained by solving the ODE in (2). Obviously, the velocity distribution is determined by the traffic rules and the traffic light states, hence the model can be generalized to other traffic scenarios easily by adjusting the velocity profile. The resulting car density profile $n(x,t)$ serves as the foundation for the network modeling later on.

To incorporate randomness into the traffic model, the deterministic variables in (1) and (2) are converted to random variables. Hence, the terms in equations (1) and (2) has the following statistical meaning,

$$c^{+}(x,t) = \frac{\partial^2 E[\mathrm{C}^{+}(x,t)]}{\partial x \partial t}, \; c^{-}(x,t) = \frac{\partial^2 E[\mathrm{C}^{-}(x,t)]}{\partial x \partial t}, \quad q(x,t) = \frac{\partial E[Q(x,t)]}{\partial t}, \text{ and } n(x,t) = \frac{\partial E[N(x,t)]}{\partial x}, \quad (3)$$

where E[•] is the expectation operator. Depending on the underlying probability model, the exact PDF for the number of cars can be found by solving either the ODE in (2) or the PDE in (1).

In the following analysis, we assume that cars arriving from the origin according to a non-homogeneous Poisson process $A$ with a non-negative and integrable external arrival rate $\alpha(t)$; then the mean number of cars arrived in the time interval $(t_1, t_2)$ can be found as $\int_{t_1}^{t_2} \alpha(t)dt$. This assumption has been verified in [17] with empirical data in Central London. Given a Poisson arrival process, a deterministic velocity profile $v(x,t)$ that is independent of $n(x,t)$, we can solve for $n(x,t)$ according to (2). Additionally, the following properties hold:

i. $N(x,t)$ and $Q(x,t)$ follow a Poisson distribution with means $E[N(x,t)] = \int_{\sigma(x,t)}^{t} \alpha(s)ds$ and $E[Q(x,t)] = \int_{-\infty}^{\sigma(x,t)} \alpha(s)ds$, respectively, where $\sigma(x,t)$ denotes the time instant at which the car arrived will remain in the road segment $(0,x]$ at time *t*;

ii. The number of cars, $N(x_1, x_2, t)$ and $N(x_3, x_4, t)$ at time $t$, in two non-overlapping regions $(x_1, x_2]$ and $(x_3, x_4]$, are independent and Poisson distributed with means

$$\bar{N}(x_1, x_2, t) = E[N(x_1, x_2, t)] = \int_{x_1}^{x_2} n(s,t)ds, \quad \bar{N}(x_3, x_4, t) = E[N(x_3, x_4, t)] = \int_{x_3}^{x_4} n(s,t)ds. \qquad (4)$$

Before moving into the network modeling, note that the independence of $v(x,t)$ and $n(x,t)$ might not hold in reality. An immediate consequence is that properties (i) and (ii) are not strictly valid. However, we still use the stochastic model as an approximation in this paper as we know from the previous result [17] that the impact of $n(x,t)$ on $v(x,t)$ is not that significant given that the traffic load is not too high (e.g., < 30 cars/min), in which case the stochastic model is still a good approximation.

To summarize, the average vehicular density is modeled by the fluid model, while the contingent variation in traffic distribution is described by the stochastic model. By providing an initial velocity profile $v(x,t)$, the dynamics of the vehicular density $n(x,t)$ is well captured by the stochastic traffic models and serve as the input to the following 802.11p protocol models. Interested readers are referred to [17] for details regarding the traffic models.

### 3.2 802.11p Protocol and Performance Models

The IEEE 802.11p standard was designed specifically for inter-vehicle communications, so that traffic safety can be enhanced through the broadcasting of safety and warning messages. The communication operates alternatively between the service channel (SCH) and the control channel (CCH) with a period of 100 ms (i.e., CCH/SCH interval + guard interval = 50 ms). The control channel is responsible for the car status beaconing (e.g., car position, speed, heading, etc.) and safety message broadcasting (e.g., any accident ahead, sudden brake, poor road condition detected, etc.). As these messages are intended to inform all nearby vehicles, broadcasting is used in the CCH. In this paper, we focus on modeling the beacon broadcasting performance of a vehicular network.

The beaconing messages are generated periodically at a rate of 10 Hz in every vehicle (node), which will be broadcasted when a car operates in the CCH. Cars are competing for the channel time according to the CSMA/CA contention mechanism. The message will be dropped if it is not sent by the end of the control channel interval as new beacon will be generated in the next channel cycle. However, according to previous studies (e.g., [29]), the dropping of beacon messages rarely happens. In addition, collided beacons will not be retransmitted as they become obsolete.

The CSMA/CA contention mechanism was modeled by a 2D Markov chain in the work of Bianchi [5]. In our case, as retransmission is not considered for beacon messages, the 2D model is reduced to a 1D model. Furthermore, to simplify the mathematical model, several assumptions are made in the MAC and PHY layers and applied to the rest of the broadcasting modeling, which are tabulated in Table 3.3.

In the following subsections, we use a Markov chain to model the transmission rate of each node in the network, from which the network performance such as broadcast success rate, delay and throughput could be estimated based on the knowledge of the vehicular density. Notations and symbols used in the derivation of the model are summarized in Table 3.4.

Table 3.3. Assumptions concerning the protocol model.

| | |
|---|---|
| PHY Layer | 1. The channel is isotropic and homogeneous across the road.<br>2. The road is seen as a one dimensional space.<br>3. Every node has the same transmission range $R_S$, and the same interference range $R_I$.<br>4. Collision happens when nodes within $R_I$ of the receiver transmit at the same time as the sender.<br>5. Collision from other nodes, including hidden nodes, is the only cause of transmission failure. |
| MAC Layer | 1. There is only beaconing message in the system queue.<br>2. There is a beacon message ready to be transported at the beginning of each CCH.<br>3. Beacon message is not affected by channel switching; hence a beacon packet treats the channel time as infinitely long [10, 30].<br>4. Beacons are generated at a rate of 10 Hz.<br>5. Since we are modeling a single queue scenario, the AIFS interval is neglected, which can be incorporated by adding extra states to the Markov chain.<br>6. Every car in the road segment is assumed to share the same MAC setting (e.g., contention window size, slot time, etc.). |

Table 3.4. Notations and symbols for the protocol model.

| | |
|---|---|
| *Notations and symbols related to network modelling (saturated queue)* | |
| $W_{ss}$ | The contention window size |
| $b_k$ | The proportion of cars in the state *k* during saturated queue modelling |
| $p(\alpha)$ | The probability that the channel is busy in the coming physical timeslot, sensed at location $\alpha$ |
| $\tau(\alpha)$ | The probability that a car at location $\alpha$ will start transmission in the coming physical timeslot |
| $R_I$ | The interfering range |
| $R_S$ | The transmission range |
| $\bar{N}(R)$ | The average number of car in range $R$ |
| *Notations and symbols related to network modelling (non-saturated queue)* | |
| $j$ | The number of virtual time slots elapsed since the beginning of the current channel time |
| $\pi_k(j)$ | The proportion of cars in state *k* at the *j*-th virtual timeslot |
| $p_g$ | The probability of generating an emergency packet in the coming virtual timeslot while a node is in idle state |
| $p(j)$ | The probability that the channel is sensed busy at the *j*-th virtual timeslot |
| $p_r$ | The probability that a new packet is generated before the message queue becomes empty |
| $\tau_{TR}$ | An estimate for the transmission probability of cars in the target range of beacon broadcasting |

### 3.2.1 802.11p Contention Model with Saturated Traffic

In saturation mode, we assume that cars sense a physical timeslot as busy with probability $p(a)$, and they transmit with probability $\tau(a)$, where $a$ is the location of the car; this assumption is similar to that of [11]. For an arbitrary node, the contention process can be characterized by a Markov chain as illustrated in Figure 3.3, with each state represents the contention counter value. The contention counter freezes in busy time slots, and counts down by one in every idle slot sensed. When the counter reaches zero, the message will be broadcasted in the next idle timeslot, and the counter will be reloaded to a random non-zero state with equal probability.

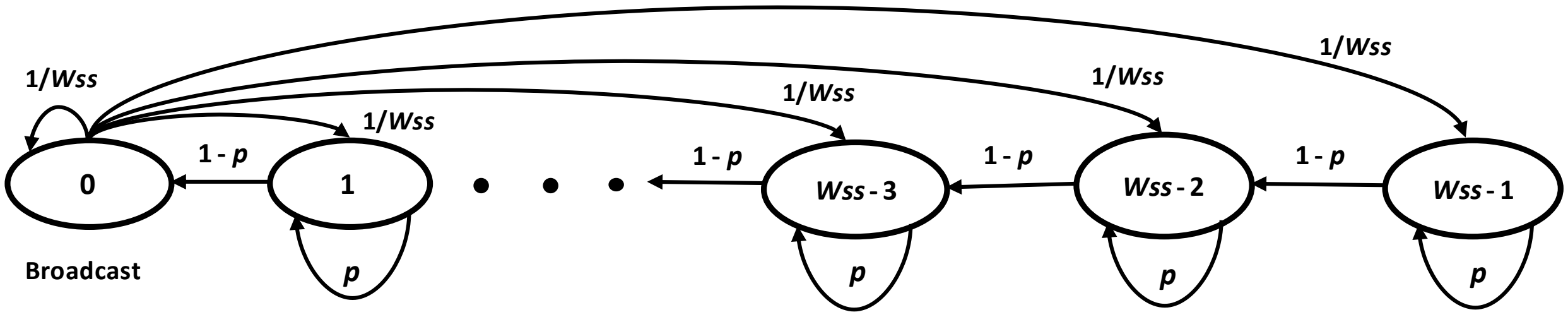

Figure 3.3. Markov chain representation of channel contention.

Exploiting the state transition relationship, we can express the transmission probability $\tau$ as,

$$\tau = b_0 = \frac{2(1-p)}{(1-2p+W_{ss})}. \quad (5)$$

At the same time, the channel busy rate $p$ can be found as

$$p = 1 - \Sigma_{k=0}^{\infty} \frac{(1-\tau)^k \bar{N}(R_I)^k e^{-\bar{N}(R_I)}}{k!} = 1 - e^{-\bar{N}(R_I)\tau}. \quad (6)$$

In the above equation, $\bar{N}(R_I)$ is the average number of cars within the sensing region. This is where the traffic model couples with the network model; the output car density profile $n(x,t)$ will be used to calculate $\bar{N}(R_I)$ based on (7) while the distribution is Poisson according to the Poisson arrival assumption.

$$\bar{N}(R_I) = \int_{R_I} n(s) ds. \quad (7)$$

The $t$ in the expression is dropped out here as the car distribution can be considered as static during a control channel interval (50 ms). Combining (5) and (6), the transmission rate $\tau$ as a function of the location space can be solved numerically according to [11]. The details of the derivation of the saturated case can be found in [18].

### 3.2.2 802.11p Contention Model with Unsaturated Traffic

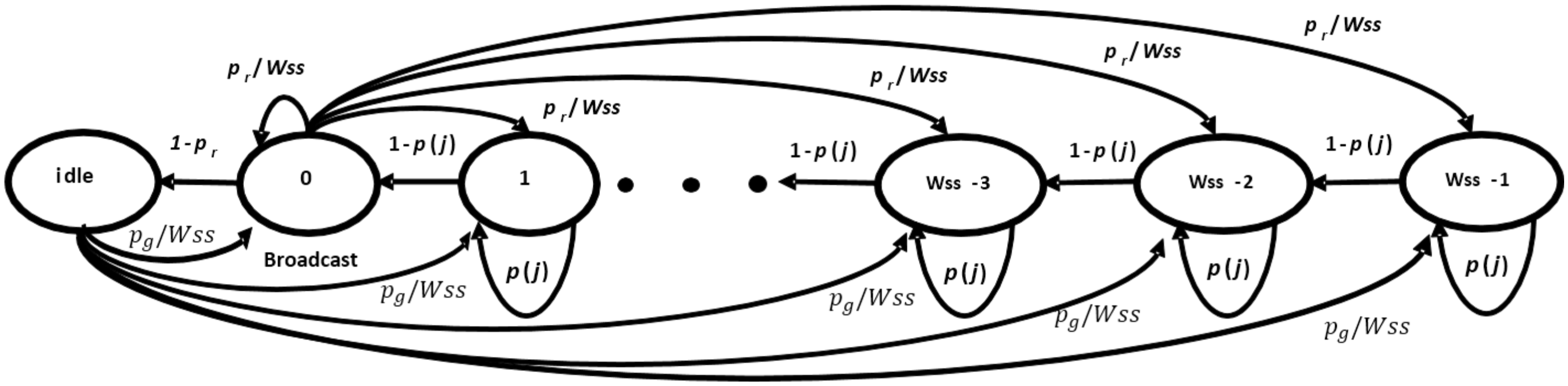

Figure 3.4. Contention representation under unsaturated traffic condition.

In the previous subsection, the message queue with saturated traffic is assumed. However, in the control channel, the majority of the traffic is beacon messages (with a generation rate of 10 Hz) for positioning. Even though there are emergency messages, they are scarcely generated. Hence, the queue is far from saturated in the control channel [31]. Therefore, we modify the model in Figure 3.3 to the one in Figure 3.4 to describe the unsaturated condition.

There are four major changes in the unsaturated state diagram: i) a new state called "idle" is added for those cars whose message queue is empty, therefore they are no longer contending; ii) the channel busy probability changes with time $j$; at the beginning of every control channel interval, the number of contending nodes is at the peak level, as every node has a beacon to send; $p(j)$ decreases as time goes by when more nodes dispatch their beacons and stay in the idle state; iii) time (*i.e.*, $j$) is measured in "virtual timeslot", whose length varies based on the channel state (*i.e.*, busy or idle); and iv) other safety related packets are generated infrequently with probability $p_g$ or $p_r$. For the definition of mathematical symbols used in the current section, please refer to Table 3.4.

In Figure 3.4, as we only consider beacon messages (*i.e.* $p_g$ and $p_r$ are essentially zero), "idle" is treated as an absorbing state in the following analysis. As the chain is no longer Markovian after this modification, the proportion of cars in each state shall be calculated recursively as

$$\begin{cases} \pi_i(1) = \frac{1}{W_{ss}} & \text{for } i = 0,1,\dots,Wss-1 \\ \pi_{idle}(j) = \pi_{idle}(j-1) + \pi_0(j-1) * (1 - p(j-1)) & \text{for } j = 2,3,\dots \\ \pi_k(j) = \pi_{k+1}(j-1) * [1 - p(j-1)] + p(j-1) * \pi_k(j-1) & \text{for } k = 0,1,\dots,W_{ss}-2 \;\; j = 2,3,\dots \\ \pi_{wss-1}(j) = \pi_{wss-1}(j-1) * p(j-1) & j = 2,3\dots \\ p(j) = 1 - e^{-\bar{N}(R_I)\pi_0(j)} & \text{for } j = 1,2,\dots \end{cases} \quad (8)$$

The first equation sets the initial state; as there must be one beacon waiting for transmission at the start of a CCH interval, nodes are uniformly distributed in each contention state at the beginning. The next three equations govern the state transition: take the expression for $\pi_k(j)$ (*i.e.* the proportion of cars in state $k$ at time $j$) as an example, it consists of the transition from upper state $\pi_{k+1}$ with transition probability $1 - p(j-1)$ and retention of the current state $\pi_k$ with probability $p(j-1)$. Both states are observed in the previous timeslot $j-1$. The last equation calculates the channel busy rate $p(j)$, which is the probability that at least one node are transmitting. $\pi_0(j)$ represents the proportion of cars whose contention counter reaches zero, and hence initiate transmission at the next timeslot (*i.e.*, $j+1$) if the channel is idle. Therefore, the transmission rate at any time slot can be expressed as $\tau(a,j) = \pi_0(j-1) * (1 - p(j-1))$, where $a$ is the location of the node. With more and more cars finish their transmission and rest in the idle state, $\pi_0(j)$ approaches 0 when $j$ is sufficiently large.

Table 3.5. Pseudo code for implementing the unsaturated contention chain.

```
Get local car number from the traffic model at the considered location a;
Set time to zero;
Set number of cars in the idle state to zero;
Initialize each state to 1/W_ss;
while (within the CCH interval and there are packets pending for transmission){
    update time slot;
    calculate the channel busy probability;
    update each contention state according to equation (8);
    update remaining contending car number;
    output the transmission probability for the current timeslot;
}
```

As the contention model can be solved by iterative calculation, Table 3.5 lists the algorithm for implementing the contention process, and the output of which is an array storing the transmission probability for every timeslot. This model produces a temporal distribution of transmission rate across the channel time, from which average transmission rate, average delay and hence throughput can be derived as shown in the following sections.

Furthermore, the simple model presented above can be easily modified to incorporate the effect of immediate access [19, 32], in which case cars can skip the contention process. This happens when the channel is sensed idle for more than AIFS before new packet arrivals. Immediate access is expected to bring a non-negligible impact on throughput in unsaturated channel after the burst of beacon broadcasting (e.g., for sparse emergency packets). As we are focusing on beacon broadcasting right now, interested readers are referred to Appendix C for details.

When calculating the Broadcast Performance Index (BPI) in the next subsection, double integral of the transmission rate is required. However, performing the algorithm at each integration step is computationally expensive, hence, an approximation for the unsaturated transmission rate is used for computing the BPI, while the average delay calculation still invokes the unsaturation model stated in (8).

### 3.2.3 Broadcasting Efficiency Modeling

Starting from this subsection, the output from the traffic model and the contention model are utilized to calculate the network performance indicators. To demonstrate the proposed joint traffic-network modeling methodology, beacon broadcasting is considered in the performance evaluation. First, we introduce the broadcast performance index (BPI). The mathematical symbols related to network performance are listed in Table 3.6.

Table 3.6. Notations and symbols used in broadcasting performance modelling.

| | |
|---|---|
| $\bar{N}_{TR}$ | The average number of cars in the beacon target range |
| $\bar{N}_{IR}$ | The average number of cars in the beacon interference range |
| $\bar{N}_a^b$ | The average number of cars in the road region $(a, b]$ |
| $\tau_{ab}$ | An estimate of the transmission probability of cars in the road region $(a, b]$ |
| $\tau_{sat}$ | Transmission probability of cars when the message queue is treated as saturated |
| $\tau_{unsat}$ | Transmission probability of cars when the message queue is treated as non-saturated. |
| $E[\text{tx time}]$ | The expect length of the transmission time |
| $\rho(a)$ | The throughput at location $a$ |
| $BPI(a)$ | The broadcasting performance index at location $a$ |
| $\xi_a$ | Broadcasting efficiency at location $a$. It is equivalent to $BPI(a)$. |
| $\xi_{a\|b,c}$ | BPI at location $a$, given effective interference sources at locations $b$ and $c$. |
| $Delay\ (a)$ | The expected delay for broadcast sent from location $a$ |

Due to potential simultaneous beacon broadcasts (failure of random backoff) and the presence of hidden nodes [33] (failure of carrier sensing), not every targeted receiver can receive the broadcast message successfully. Previous works [10, 11, 30, 34, 27, 7, 12, 13, 29] defined a broadcast failure when at least one of the targeted receivers get interfered (*a.k.a*, success or failure); this overlooks the fact that certain number of cars can still receive the broadcast message despite others being corrupted. As we are investigating the spatial distribution along the road, we need a metric that takes into account the partial failure/success nature of a broadcast. As a result, the broadcasting performance index (BPI) at location $a$ is defined as follows,

$$\xi_a = \frac{\text{\# of interference−free targeted nodes}}{\text{\# of targeted nodes}}. \quad (9)$$

Assume that only cars in the backward direction of the transmitter are interested in the broadcast message (since drivers usually want to know the preceding traffic rather than the traffic at the back), thus the set of target audiences for the broadcast message from a car at location $a$ are those within its 'backward' transmission range ($R_s$). Among those targeted receivers, a portion of them may be interfered by other concurrent transmitters in the network; for example, cars at locations $b$ and $c$ as illustrated in Figure 3.5. Depending on the location $b$ and $c$, they could be a hidden node to the sender $a$. For brevity, we use $a$, $b$ and $c$ to denote the nodes as well as their respective locations in the following analysis. Based on the definition in (9), the BPI at location $a$, given the effective interference sources at locations $b$ and $c$, can be expressed as

$$\xi_{a|b,c} = \left(1 - \frac{\bar{N}_{IR}}{\bar{N}_{TR}}\right) * e^{-\tau_{TR} * \bar{N}_{TR}}, \quad (10)$$

where $\bar{N}_{IR}$ and $\bar{N}_{TR}$ are the expected number of cars within the interfered region (IR) and targeted region (TR) as illustrated in Figure 3.5 respectively, which can be calculated from (2) and (3). $\tau_{TR}$ is the estimated transmission rate of cars within the targeted region that can be obtained from (8). The effective interference sources are defined as the closest cars that transmit simultaneously with node $a$. Due to the fixed transmission range defined, those cars located beyond the effective interference sources cause no effect on the BPI.

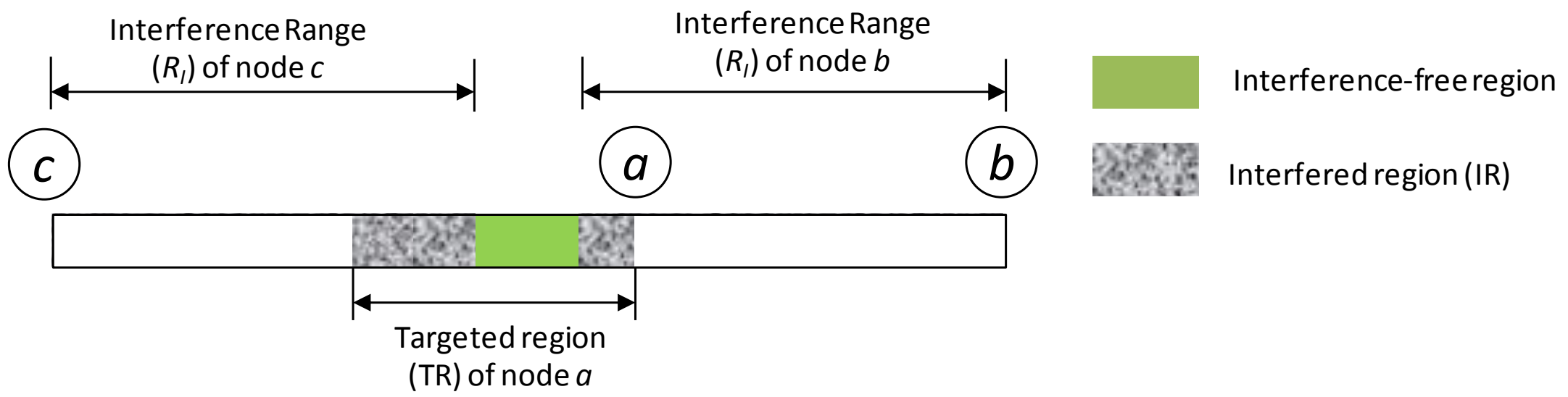

Figure 3.5. Illustration of the targeted region (TR) and interfered region (IR).

Lastly, the overall BPI at location $a$ can be computed as follows,

$$\xi_a = \int_{-\infty}^{a-Rs} \int_{a}^{+\infty} \xi_{a|b,c} * P(B = b) * P(C = c) db dc, \quad (11)$$

where $Rs$ is the backward transmission range of node $a$, and $P(B = b)$ is the probability that an effective interference source locates at $b$ which can be expressed as

$$P(B = b) = \tau_{ab} n(b) e^{-\bar{N}_a^b \tau_{ab}}. \quad (12)$$

In (12), $\bar{N}_a^b$ is the average number of cars within $(a, b]$. $\tau_{ab}$ is the average transmission rate within the same region, which can be calculated by (8) for the beacon message case. $n(b)$ is the car density at location $b$. $P(C = c)$ can be obtained in the same manner. It is assumed that the effective interference sources $b$ and $c$ are located beyond the targeted region or backward transmission range of node $a$, since otherwise $\xi_{a|b,c} = 0$.

As mentioned at the end of Subsection 3.2.2, the calculation of the unsaturated transmission rate via (8) becomes computationally expensive as each integration step in (11) requires a run of the algorithm listed in Table 3.5. Therefore, an approximation of the unsaturated transmission rate from the saturated case is used in the BPI analytical evaluation and its accuracy is verified against simulation results. The following relationship was used to associate $\tau_{sat}$ with $\tau_{unsat}$.

$$\tau_{unsat} = [k_1 \log(n) + k_2(W_{ss}) + k_3]\tau_{sat}, \tag{13}$$

where $n$ is the car density at the current location, $W_{ss}$ is the contention window size, $k_1$, $k_2$ and $k_3$ are coefficients determined by iterative curve fitting for possible range of $W_{ss}$ and $n$. For saturated queues, (5) – (7) can be used directly. The reader is referred to [18] for the derivations of (5) – (7).

It is worth noting that the calculation of BPI takes two transmission failure mechanisms into consideration: i) the hidden nodes are considered by effective interfering sources located beyond the sender's sensing range; and ii) the potential concurrent transmission of nodes due to random backoff. For other failure mechanisms such as bit error, they can be incorporated into the model easily by adjusting the coefficient $\xi_{a|b,c}$. Similarly, we could also adopt a more elaborated signal propagation model for the calculation of $\xi_{a|b,c}$ and incorporate such as the channel capturing effect to characterize the case that the receiver might be able to decode the message in the presence of interference given a sufficient signal-to-interference-and-noise ratio (SINR). Interested readers are referred to Appendix D for a detailed description.

### 3.2.4 Delay Modeling

Delay here refers to the time between packet generation and reception, which comprises of the waiting time due to packet contention and the actual transmission time. For a fixed beacon size, the transmission time is constant. The total delay hence can be calculated by

$$Delay = E[\text{tx time}] + \Sigma_{k=0}^{+\infty} E[\text{waiting time}, k] * P(\text{a node starts sending in the } k+1\text{th virtual time slot}) \tag{14}$$

In the above equation, virtual time slot that depends on the transmission time is used to differentiate from the 16 μs physical time slot defined in 802.11p. $E[\text{waiting time}, k]$ is the expected waiting time for a node before it will transmit in the $k$+1-th virtual time slot. It is expressed as,

$$E[\text{waiting time}, k] = \begin{cases} \Sigma_{j=1}^{k} \{E[\text{tx time}] * p(j) + [1 - p(j)]\} & \text{for unsaturated queue} \\ k * \{E[\text{tx time}] * p + (1 - p)\} & \text{for saturated queue} \end{cases}, \tag{15}$$

where $E[\text{tx time}]$ is the expected length of the virtual time slot when the channel is busy, and the busy state remains until the transmission completes (e.g., for transmitting a packet of 500 bytes over a 3 Mbps channel, the length of a virtual time slot ≈ 80 physical time slots = 80 * 16 μs).

### 3.2.5 Receiver-end Broadcasting Throughput Modeling

The broadcasting throughput characterizes the broadcasting capability at different locations in the network under the CSMA/CA mechanism. The receiver-end broadcasting throughput at location $a$ is defined as the number of successfully received broadcast packets sent from a car located at $a$ per unit time. Generally, a broadcast is considered as successful only if the message reaches all the targeted receivers. Since we have introduced the concept of broadcasting performance index (BPI) in Section 3.2.3, we study the broadcasting throughput here by considering the number of targeted receivers that successfully receive the broadcast message. Hence, we have

$$\begin{aligned} \rho(a) &= \#\text{ of received packets/Delay} \\ &= E[\text{Targeted Cars}] * BPI(a)/Delay\ (a) \\ &= \bar{N}_{TR} * BPI(a)/Delay(a), \end{aligned} \quad (16)$$

where $\bar{N}_{TR}$ is the expected number of audiences in the targeted region (TR) of node $a$, obtained by integrating the density profile from the stochastic traffic models over the backward transmission range $R_S$ of node $a$. The BPI and delay at location $a$ can be found respectively by (11) and (14).

# 4. Numerical Results with Vehicle Interaction

In this section, simulation is conducted to verify the proposed traffic-network model on beacon broadcasting. Note that in deriving (2), the independence of velocity $v(x,t)$ from traffic density $n(x,t)$ was assumed, which allowed (2) to be solved as an ODE. However, when vehicle interaction is introduced into the system through the density dependent velocity profile, the stochastic independence and Poisson distribution assumed in the analytical models in principle do not hold, and the derived analytical results can only serve as an approximation. We exploit simulation in this section to validate the quality of such approximation for respective performance indexes.

## 4.1 Simulation Scenario

A C++ program was developed to simulate the 802.11p vehicular network in a one-dimensional road setting. The road was semi-infinite with cars arriving from the left according to a Poisson arrival process as illustrated in Figure 3.1. Cars entered the road segment and move with a mean free speed of 1 km/min, and car-to-car interactions were taken into account by incorporating the revised Greenshield's model (17) into simulation. In (17), $v_f$ is the mean free speed, and $k_j$ is the jamming density.

$$v(x,t) = v_f(1 - \frac{n(x+\Delta x,t)}{k_j}), \quad (17)$$

The first simulated scenario has homogeneous car distribution along the road at various density levels, with the intention to reveal the general network performance at different levels of vehicular traffic load. The second simulated scenario has a traffic light located at 2 km from the starting point, which was designed to test the model accuracy under non-homogeneous traffic conditions. The traffic light turned red for time $t \in (4,4.5]$ min, while the light remained green at other time. During the green light duration, the mean free speed $v_f$ remains constant, while it follows the spatial distribution depicted in Figure 4.1.1 during the red light duration.

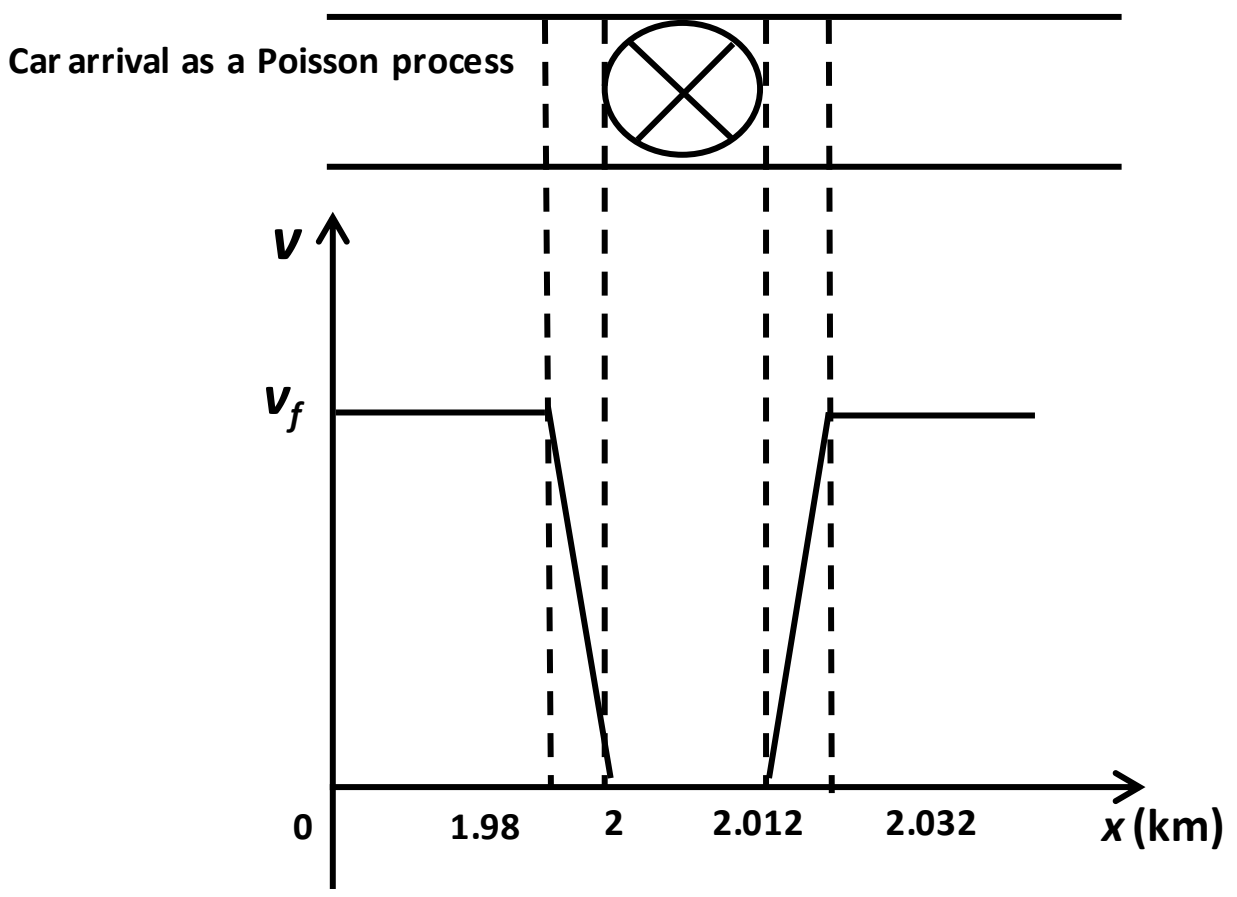

Figure 4.1.1. The mean free speed as a function of location for the traffic light scenario during the red light duration.

In addition, channel homogeneity is assumed, that is, every vehicle has the same transmission range, interference range, etc. as listed in Table 4.1.1. For every simulated vehicle, we assume that one beacon packet is loaded to the message queue for broadcasting at the beginning of every CCH interval [26, 35, 7, 29, 13, 11].

Table 4.1.1. Network parameters used in the simulation.

| **Parameter** | **Value** | **Parameter** | **Value** |
|---|---|---|---|
| Transmission Range ($R_S$) | 0.2 km | Packet Size | 500 bytes |
| Interference Range ($R_I$) | 0.5 km | Data Rate | 3 Mbps |
| Maximum Contention Window Size ($W_{ss}$) | 4, 8, 16, 32, 64 | Beacon Generation Rate | 10 Hz |
| Slot Time | 16 µs | Channel Period (CCH + SCH) | 100 ms |

## 4.2 BPI Performance

Figure 4.2.1 shows the BPI result for the homogeneous traffic scenario. The proportion of successfully received packets was recorded for each car instance in every control channel cycle, and the final result was obtained by averaging the BPI of all car instances along the road and the total number of simulation trials.

The goodness of fit of the analytical results against the simulation results is validated through the Kolmogorov-Smirnov (K-S) test [36] (with a significance level of 10%) so that the accuracy of the analytical models can be quantified. Table A.1 in Appendix A shows the relative error between the two and the corresponding row-wise and column-wise K-S test results. If the null hypothesis cannot be rejected, the result is 0 and the corresponding *p*-value is above the significance level.

In Figure 4.2.1, the BPI decreases as the traffic density increases, indicating a greater packet collision rate and hence a larger broadcasting lost in a busier channel. In all of these tests, the BPIs drop below 0.3 when the car density exceeds 30 cars/km, meaning that two thirds of the packets cannot reach their intended receivers due to collisions. Comparing the results among different contention window sizes leads to a conclusion that larger $W_{ss}$ always results in higher BPI, given the same car density.

The analytical result matches with the simulation result in general. It is worth noting that we have approximated the analytical result of BPI for the unsaturated communication traffic with (13) so as to reduce the computation cost during model evaluation. The accuracy of the analytical results can be further boosted by considering the recursive process as described in (8).

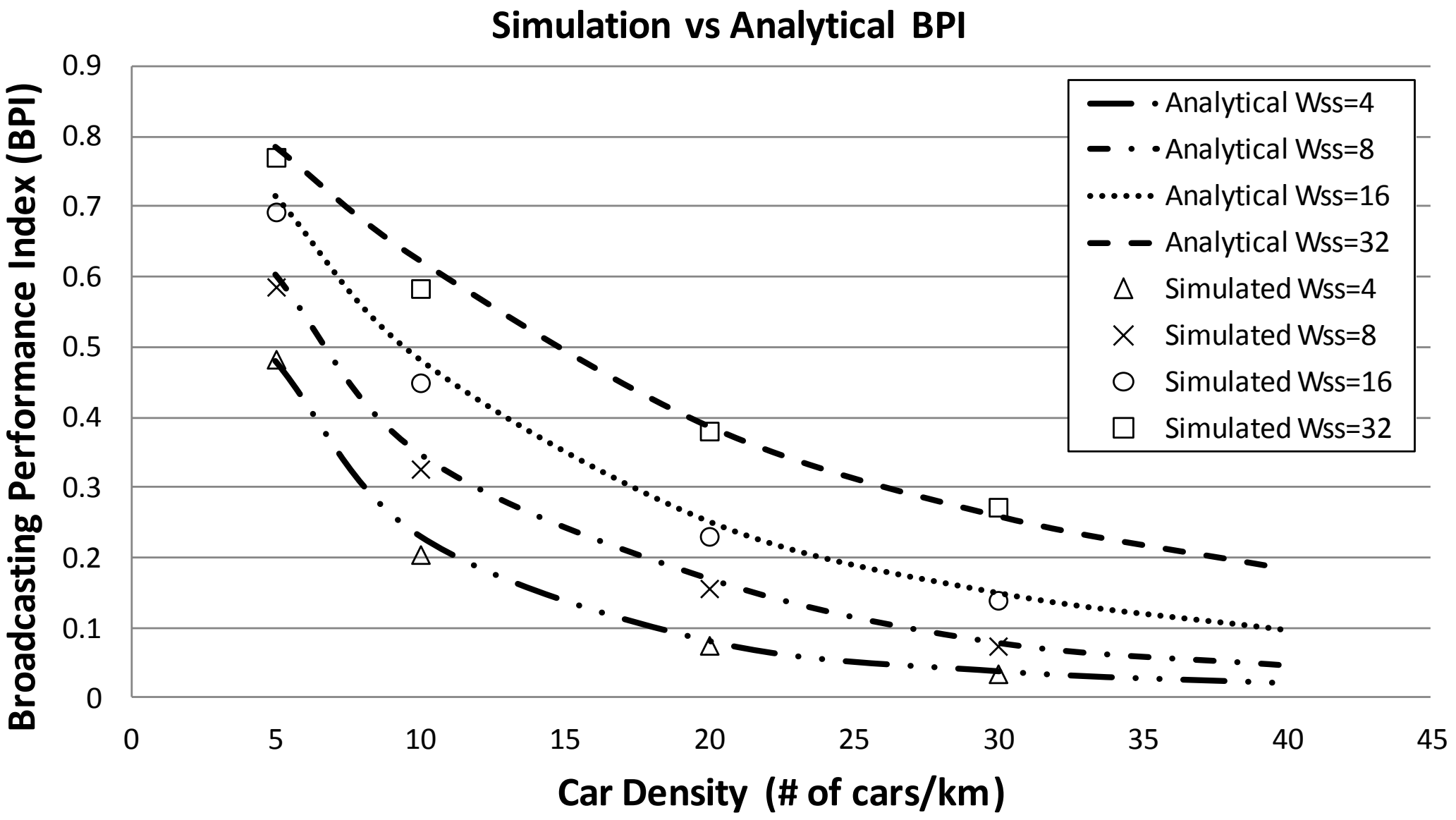

Figure 4.2.1. Broadcasting performance index (BPI) against car density for homogeneous traffic.

In the non-homogeneous case, a traffic light is introduced at location 2 km, which gives rise to the formation of a vehicular platoon during the red signal period for $t \in (4,4.5]$. Figure 4.2.2 shows the BPI performance for car arrival rate of 10 cars/min at time instant 4.5 min when the traffic light was about to turn green.

Table 4.2.1 provides a quantitative evaluation on the accuracy of the analytical model. From the table, it is quite satisfying that the average relative error between the simulation and analytical results is generally less than 10%. According to Figure 4.2.2 again, we can observe the density pulse at location 2 km, indicating that cars stop by the road junction during the red signal period, and that region reaches the jamming density. In region [2, 2.5] km, the car density drops to zero as previous cars have moved past the junction. In other regions, the car density remains roughly constant at about 10 cars/km. Two major observations can be made based on the non-homogeneous BPI result are: i) the BPI drops starting from location 1.5 km since the high traffic density ahead at 2 km hinders successful broadcasts (the 0.5 km gap is exactly the interference range as shown in Table 4.1.1); and ii) the BPI surges right after the road junction starting from location 2.5 km due to the absence of interfering cars at the road junction at this time instant.

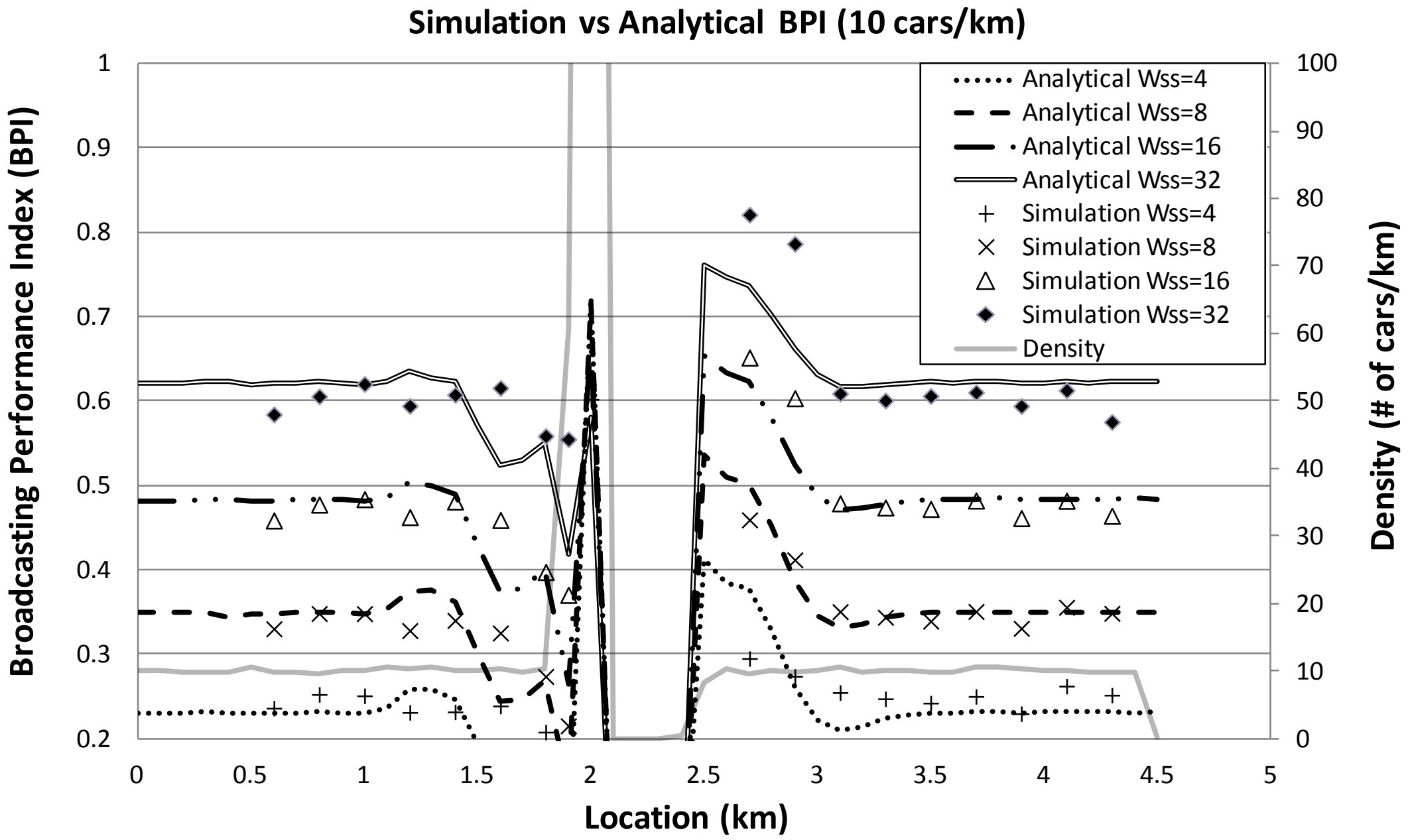

Figure 4.2.2. The simulated and analytical BPI with traffic light and non-homogeneous node distribution (λ = 10 cars/min).

Table 4.2.1. Difference between the simulated and analytical BPI with non-homogeneous node distribution (λ = 10 cars/min).

| | ***Wss* = 4** | ***Wss* = 8** | ***Wss* = 16** | ***Wss* = 32** |
|---|---|---|---|---|
| **avg. relative difference in %** | 14.71 | 7.34 | 9.98 | 11.27 |
| **avg. absolute difference** | 0.03 | 0.02 | 0.05 | 0.08 |

## 4.3 Delay Performance

Delay is defined as the average number of physical time slots required for a generated packet to reach the destination. The simulation result was obtained by averaging the number of timeslots between the start of the CCH cycle and the completion of transmission at a specific location across all simulation trials. Figure 4.3.1 plots the analytical and simulation results for the delay performance under homogeneous car distribution. The model accuracy is verified by the K-S test, and the results are tabulated in Table A.2 in Appendix A.

Generally speaking, the delay time increases as car density rises, because each car needs to compete with more neighbors for the channel access. Similarly, the delay is larger for larger contention window sizes due to longer waiting time on average.

One should note that the delay considered here is a statistical average that a randomly selected node in the network would encounter before completing a beacon transmission. Given that the control channel cycle consists of 3125 time slots (i.e., 50 ms/16 μs), it can well contain an average delay of 1200 time slots as shown in Figure 4.3.1, which confirms the assumption that a beacon message is rarely dropped by the end of a control channel cycle.

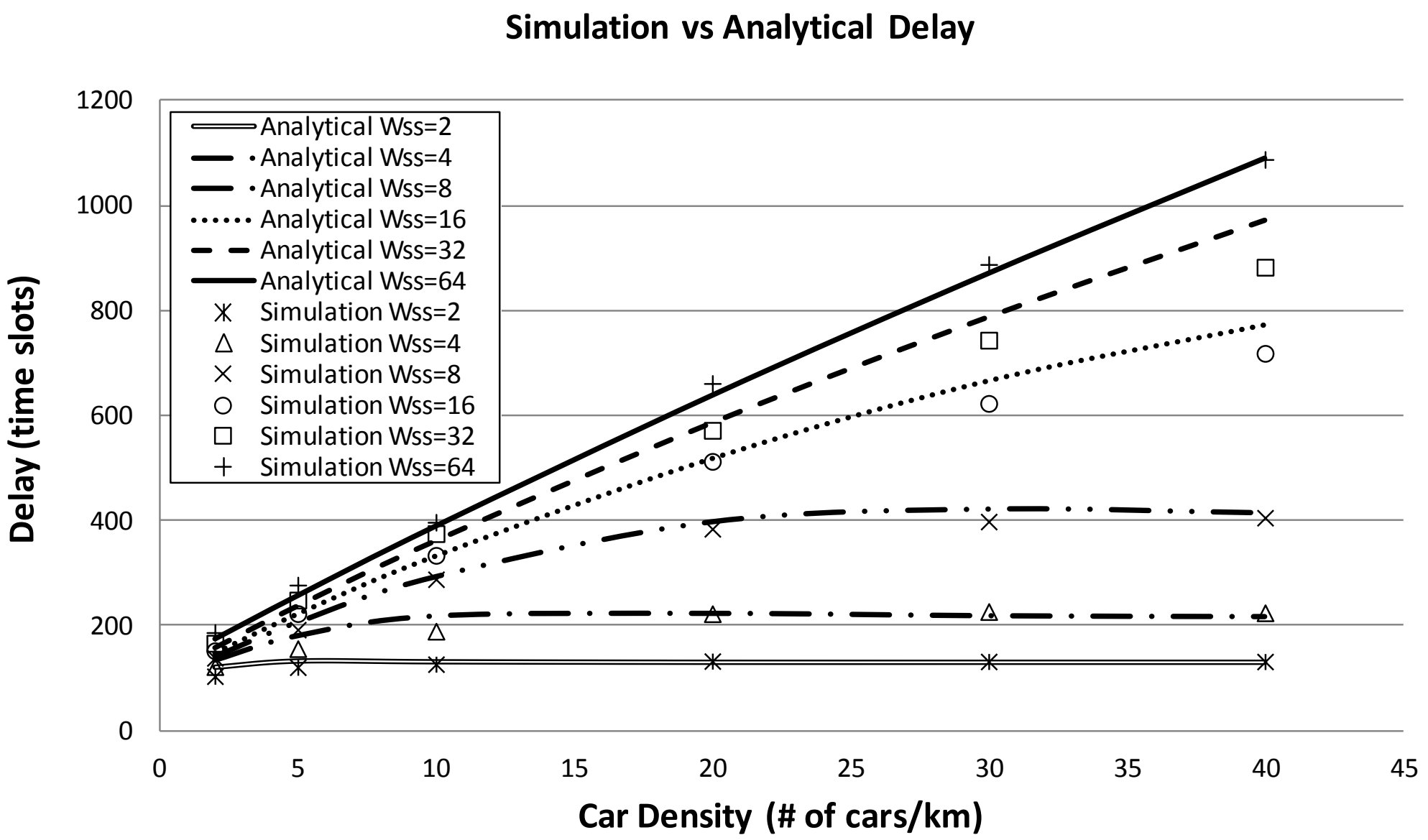

Figure 4.3.1. Delay against car density for homogeneous traffic.

According to Table A.2, the relative differences between the simulation and analytical models are less than 10% for almost all *Wss* values. In addition, the row-wise and column-wise K-S tests were performed to test the model's accuracy, and the majority of the results accept the null hypothesis with 99% confidence level.

Figure 4.3.2 shows the delay under the non-homogeneous traffic scenario, which is similar to the one in Section 4.2. By the end of the red signal period, cars accumulate in front of the traffic light, resulting in a vehicular platoon. In contrast to the BPI distribution in the same scenario (Figure 4.2.2), the delay peaks at location 1.5 km. The high traffic density ahead leads to intensive contention and long delay for cars at location 1.5 km. A similar reason explains the low delay observed at location 2.5 km due to the density gap in the region (2,2.5]. Table 4.3.1 shows the average relative error between the analytical and simulated delay, which are less than 8%.

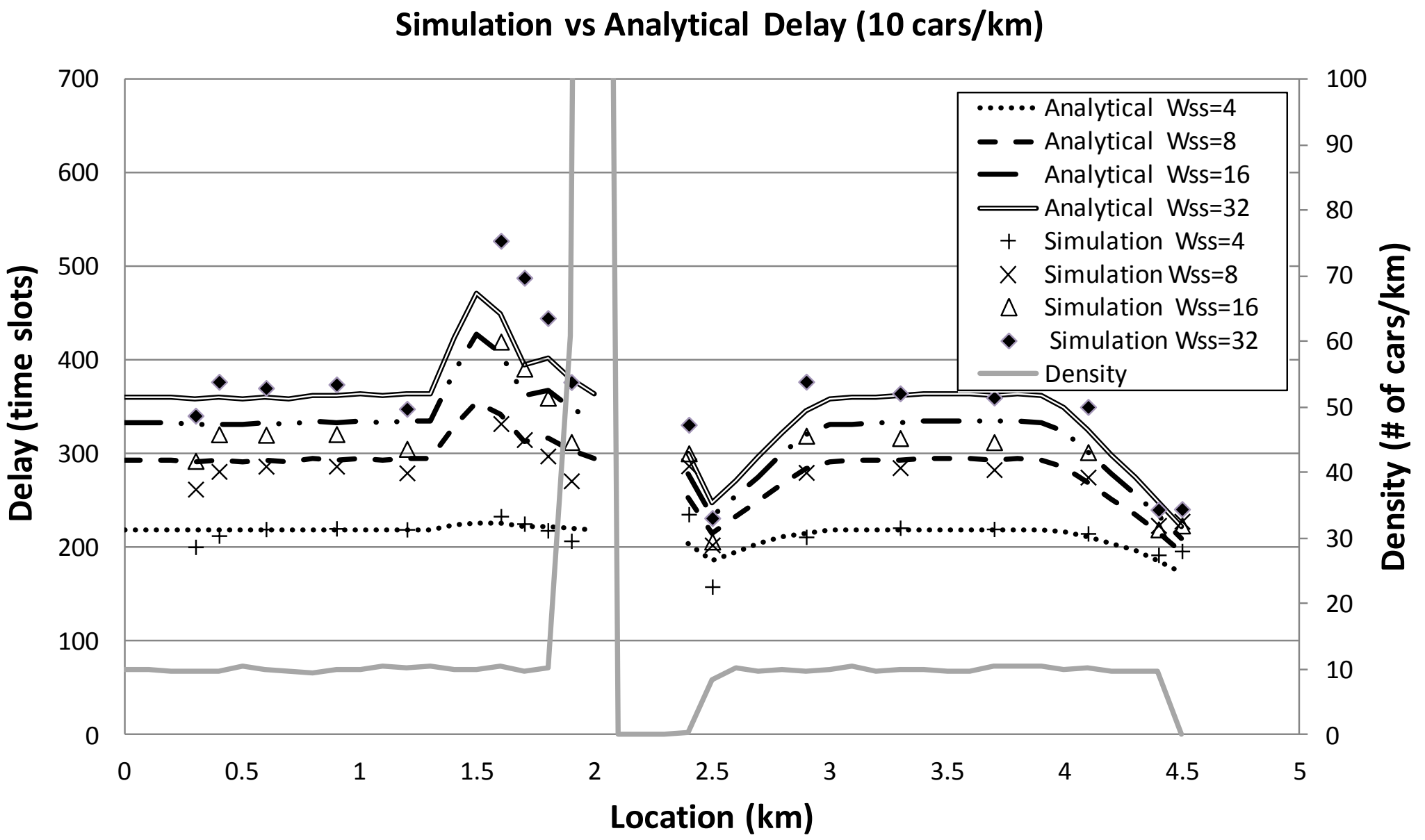

Figure 4.3.2. The simulated and analytical delay with traffic light and non-homogeneous node distribution (λ = 10 cars/min).

Table 4.3.1. Difference between the simulated and analytical delay with non-homogeneous node distribution (λ = 10 cars/min).

| | ***Wss* = 4** | ***Wss* = 8** | ***Wss* = 16** | ***Wss* = 32** |
|---|---|---|---|---|
| **avg. relative difference in %** | 2.32 | 1.28 | 7.59 | 0.79 |
| **avg. absolute difference** | 5.45 | 3.11 | 19.87 | 3.74 |

## 4.4 Broadcasting Throughput Performance

The broadcasting throughput curve in Figure 4.4.1 is obtained by extracting the total number of successfully received beacon packets in the target range of a sender divided by the transmission delay. The test results for the throughput modeling accuracy are shown in Table A.3 of Appendix A, which shows the corresponding relative error between the simulated and analytical broadcasting throughput and the K-S test results.

Unlike the monotonic trend of BPI or delay against car density, the throughput peaks at a density of about 5 cars/km for a given *Wss*, which is due to the trade-off between the number of targeted receivers and the BPI. Beyond 5 cars/km, the degradation of BPI out weights the increases in the number of targeted receivers within the transmission range, resulting in a decline in the throughput.

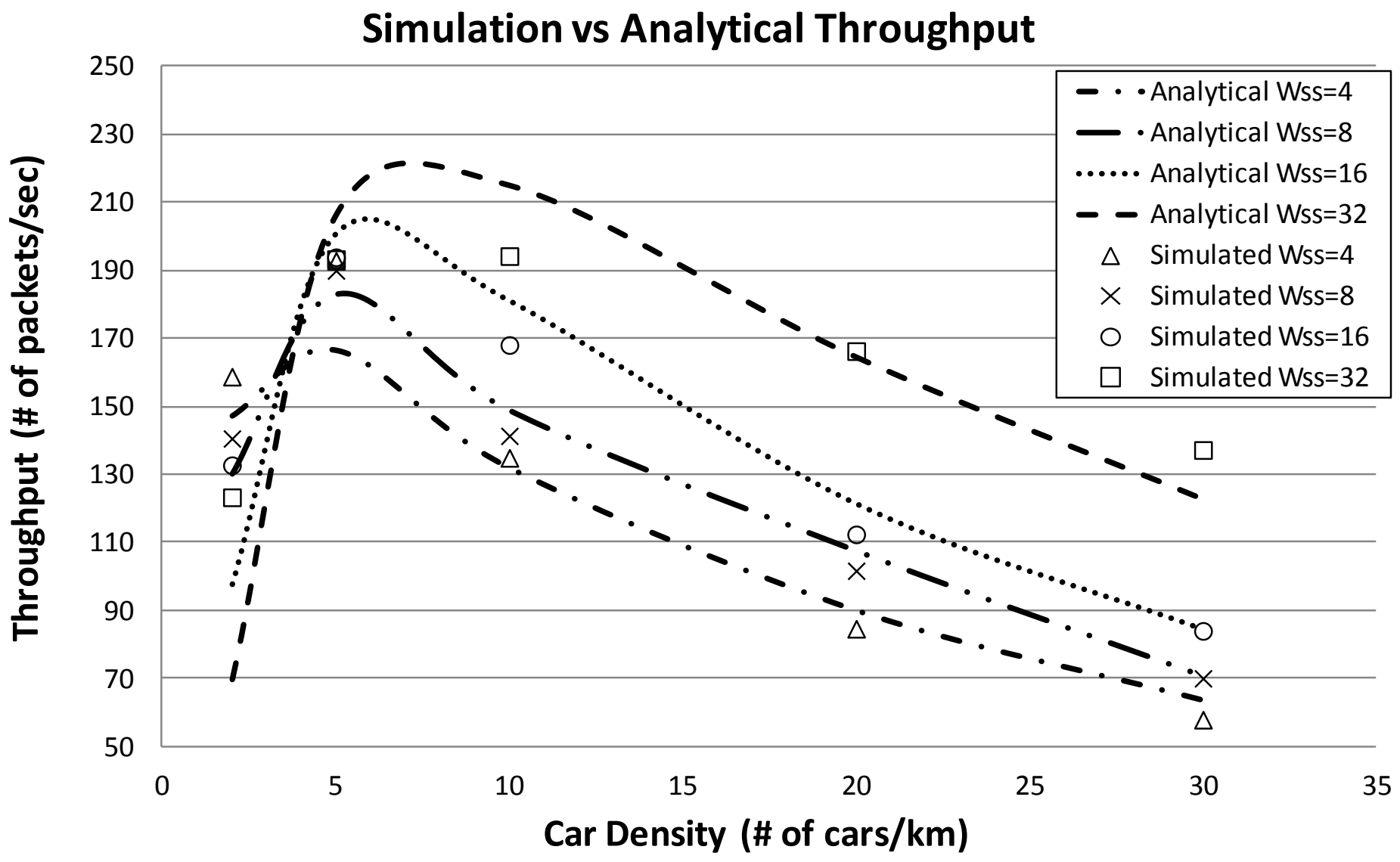

Figure 4.4.1. Receiver-end broadcasting throughput against car density for homogeneous traffic.

Figure 4.4.2 shows how the receiver-end broadcasting throughput varies with respect to the location change under non-homogeneous car distribution. The throughput decreases at location 1.5 km, following similar pattern as the BPI in Figure 4.2.2. The same is true for the region immediately after the road junction at location 2.5 km. The increase in throughput at the end of the road segment (beyond location 4.5 km) is due to the boundary effect, as no cars are considered beyond the boundary, hence no interference exists from ahead. Table 4.4.1 tabulates the average differences between the simulation and analytical results. The average relative error is generally less than 10%.

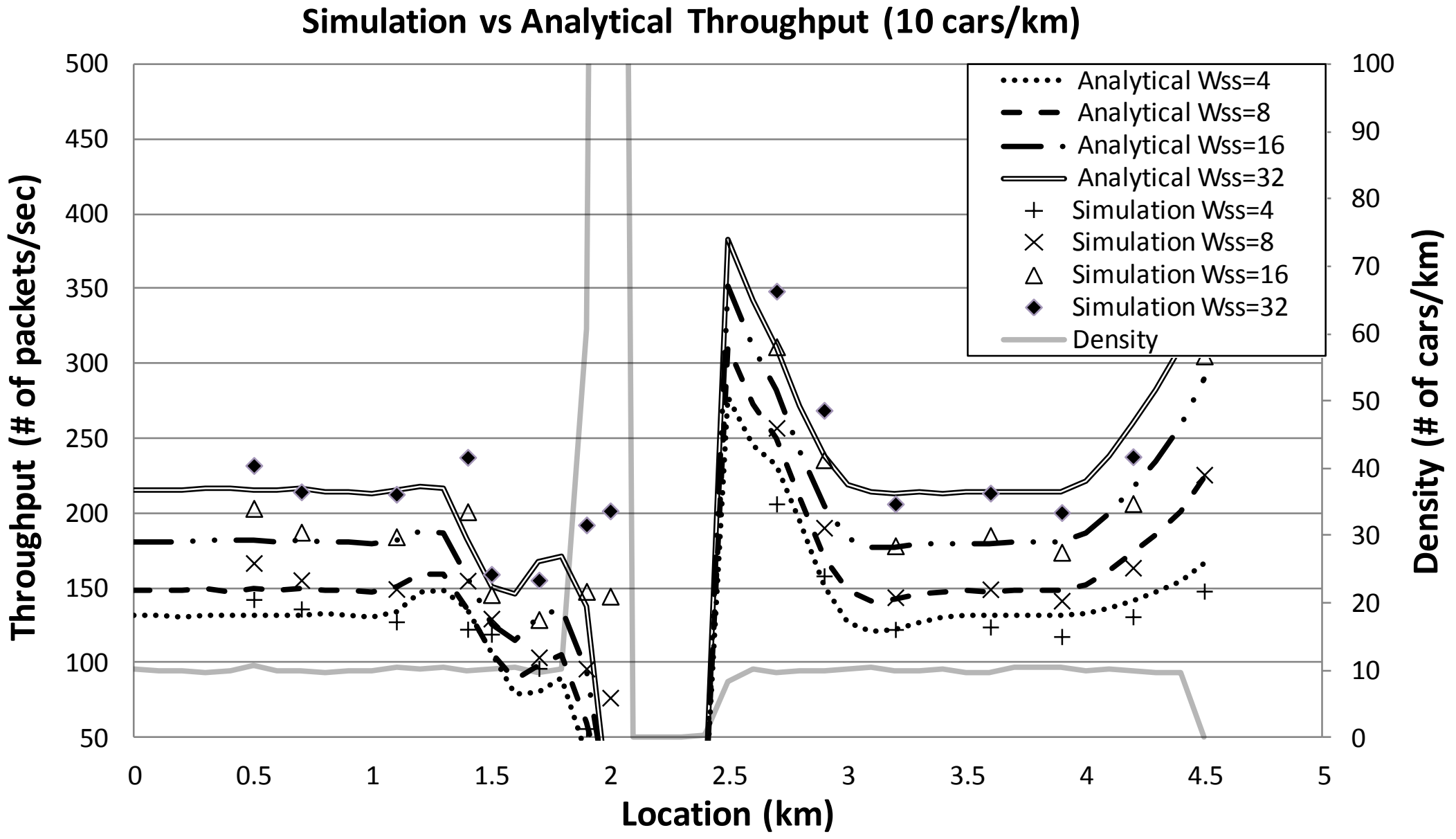

Figure 4.4.2. The simulated and analytical throughput with traffic light and non-homogeneous node distribution ($\lambda$ = 10 cars/min).

Table 4.4.1. Difference between the simulated and analytical throughput with non-homogeneous node distribution (λ = 10 cars/min).

| | *Wss* = 4 | *Wss* = 8 | *Wss* = 16 | *Wss* = 32 |
|---|---|---|---|---|
| **avg. relative difference in %** | 11.59 | 5.31 | 6.58 | 3.31 |
| **avg. absolute difference** | 0.65 | 12.44 | 21.96 | 22.21 |

## 5. DISCUSSION

The homogeneous traffic simulation and analytical results, as depicted in Figure 4.2.1, Figure 4.3.1, and Figure 4.4.1, reveal a delicate relationship among BPI, delay, throughput, car density $(n)$ and the contention window size $(W_{ss})$. The impact of $n$ and $W_{ss}$ on BPI and delay is straight forward. Higher car density brings more contenders for the channel access, raising the collision probability (hence lower BPI) and longer delay. Larger contention window $(W_{ss})$, on the other hand, makes nodes less aggressive in attempting to transmit, thus collision is less frequent (higher BPI) and the delay is longer. Finally, the throughput curves in Figure 4.4.2 exemplify the intriguing trade-off between BPI and delay: i) at low car density $(n < 5\ \text{cars/km})$, delay is the predominant factor in throughput performance (the BPI gain from larger contention window size cannot compensate for the delay loss), resulting in lower throughput for larger $W_{ss}$; ii) for medium to high car density $(n > 5\ \text{cars/km})$, BPI becomes the dominating component as the BPI gain from larger $W_{ss}$ lifts the overall throughput performance despite the deteriorated delay; and iii) the cross-over point of the throughput curve family at around 5 cars/km indicates the operating density at which the influence of BPI surpasses that of delay on throughput.

The above analysis provides a key message to strategic optimization of the beacon broadcasting throughput. That is, at low car density, we should focus on reducing delay by adjusting the contention window to small values, while at high car density, we should focus on improving BPI by adopting large contention window size.

From the safety point of view, both high BPI and low delay are desired. In Figure 4.3.1, we can observe that even for high car density $(n = 50\ \text{cars/km})$ and large contention window size $(W_{ss} = 64)$, the average delay is only about 1000 timeslots (*i.e.*, 16 ms), which is acceptable. However, BPI is the main concern. Figure 4.2.1 reveals that at medium car density, say $n = 30\ \text{cars/km}$, the BPI for large $W_{ss}$ drops below 0.3, which is far below the 99% reception success rate considered in [37] for safety applications. Therefore, effort shall be made to boost the BPI to a reasonable level for ensuring various safety operations.

To ensure high BPI, one could adjust the communication parameters adaptively. Firstly, the contention window size can be dynamically adjusted with respect to the vehicular traffic density (e.g., larger when high traffic density is encountered). Hsu *et al.* [38] investigated a contention window size offset scheme to reduce collision based on the local estimation of car density around. Moreover, the transmission power of nodes (vehicles) can be adjusted according to the traffic density as well to optimize the BPI; for example, when the car density is high, reducing the transmission range of nodes will reduce the amount of interference and mitigate the hidden node corruption. Finally, the data rate of transmission can also be exploited to improve BPI by reducing the transmission time for beacon packets, and hence lowering the chance of collision. The effectiveness of data rate control on BPI can be inferred from the comparison of the transmission success rate between [29, 13, 35] and [26, 7, 10]: the first group of references utilized a 3 Mbps transmission rate, and they generally had a lower successful delivery rate than those in the latter group, which employed a 6 Mbps rate.

Another way for improving BPI involves altering the behavior of the broadcast, such as repetitious broadcasting. The rationale behind is that if a message is repeatedly broadcasted multiple times, the proportion of nodes that receive at least one beacon increases. For example, given the single broadcast BPI as 50% and the broadcast repeats for 7 times, the probability that a targeted node successfully receives the message becomes 99% ($1 - 0.5^7 = 0.992$). Ma *et al.* [34] proposed a receiver-oriented repetition broadcast scheme for emergency message broadcast. However, the cost for repetitious broadcasting is of course information duplication.

A more aggressive approach to ensure high BPI via changing the broadcasting behavior is the centralized scheduling broadcast scheme. In such a scheme, every node is assigned with a time frame to transmit, and collision can be avoided to a great extent (except for the problem of hidden nodes). The price to pay is the coordination overhead, but the reward is high BPI. Furthermore, according to the estimation presented in Appendix B, a coordinated system has a higher throughput than the best performance shown in Figure 4.4.2. Some researchers have been studying the possible use of coordinated communication in vehicular environment. For instance, Bilstrup *et al.* proposed a self-organized time division multiple access (STDMA) approach [39], which allocates dedicated time frame for individual broadcasting node. Similarly, the multiple class scheme proposed in [34] classifies safety messages into three priority levels, and suppresses hidden node interference by a long range busy tone in another frequency channel, which is again some sort of coordination.

Whether by adjusting communication parameters or selecting an appropriate coordination scheme, there is a need for mode or parameter switching, which involves certain amount of configuration delay. Therefore, knowledge about car density distribution along the road enables individual car to decide the ideal operating point at each location in advance, and hence perform preemptive configuration to ensure a smooth state transition. Our joint traffic-network modeling methodology allows dynamic spatial network performance analysis in a road system with non-homogeneous car distribution. With the input of traffic inflow/outflow data and velocity profile, our model is able to predict the temporal and spatial variations of the network performance in a signalized urban road network. The performance evaluation can be done in a remote central server, and the result would be downloaded to each car in the region; based on the result, cars are able to adjust the communication parameters at corresponding locations.

The accuracy of the proposed methodology in predicting temporal and spatial distribution of network performance is verified against extensive simulation, the results of which are shown in Figure 4.2.2, Figure 4.3.2, and Figure 4.4.2. Based on the homogeneous scenario discussed above, the performance tends to deteriorate under high car density as expected. Furthermore, our model prediction matches the expectation that the network performance (BPI, delay and throughput) declines at 1.5 km where the cars' sensing range covers the entire traffic platoon built up in front of the traffic light; the reverse is true for the performance boost at location 2.5 km, where car density is zero in region (2, 2.5].

Spatial prediction of network performance can help the system better plan its operation. For example, if we want to maintain a BPI of 0.5 in the scenario depicted in Figure 4.2.2, given the default $Wss = 16$, then the system shall switch to $Wss = 32$ at location 1.5 km. Because there is some delay in switching the $Wss$ configuration, knowing the network performance in advance helps initiate the parameter switching routine well before the car moving into the traffic jam area. The same mechanism can assist communication mode switching. Every car within a region shall obtain the same instructions given the same set of input; therefore, before entering the high traffic density zone, every car would switch to the coordinated broadcasting mode so as to realize preemptive mode switching. The ability to capture the spatial and temporal network performance variations on a realistic traffic distribution enables the proposed models to support preemptive communication optimizations; these are not feasible with conventional simulation approaches, which are too time consuming to provide real time prediction.

Overall, we have coupled the non-saturated network queue model with the traffic model and demonstrated that the proposed set of models can successfully estimate the non-homogeneous network performance distribution in this paper, which lays the foundation for future implementation of preemptive network optimizations. Future extension of the current work includes the following:

1. To incorporate other message types besides beacon broadcasting. An immediate consequence is the deterioration of BPI and delay. For example, in Figure 4.3.1, the delay values for low $Wss$ level off when car density increases beyond a threshold. This is because we assumed that there was only one beacon to send and it was ready at the beginning of each control channel cycle; under small $Wss$, nearly all cars chose to send the beacon at the beginning of the channel time and entered an idle state, making it easier for other contending nodes to access the channel and bringing a cap to the delay value. If we considered additional messages, it is expected that the delay will continue to grow with car density.

2. To incorporate other realistic packet delivery failure modes. Currently, as a demonstration of the joint network-traffic modelling platform, we want to keep the model simple and essential, and only considered packet delivery failure caused by hidden nodes and concurrent transmissions since they are closely related to car distribution. Other realistic failure modes, such as fading in the communication channel can be added to the model by probabilistically modifying certain terms.
3. To incorporate the EDCA mechanism. A more realistic 802.11p network should include multiple access categories (ACs), but due to the space limit, the derivation of transmission rate for individual AC queue after internal contention is not presented here. Once we obtain the transmission rate of each AC queue, we can calculate the BPI, delay and throughput for each queue accordingly using (10), (14) and (16).

## 6. CONCLUSIONS

In this paper, a joint transport and communication modeling methodology is proposed to evaluate the broadcasting performance of IEEE 802.11p in a realistic signalized urban road segment. The integrated approach from characterizing the vehicular traffic dynamics to the modeling of various network performance indexes highlights the importance of the consideration of non-homogeneous car distribution and the association between the traffic and network models on 802.11p VANET performance evaluation.

The traffic models not only capture the macroscopic property of a traffic flow, but also take into account the non-homogeneity of node distribution due to car-to-car interaction and traffic control. Additionally, the stochastic dynamics and randomness of the vehicular traffic are also incorporated. By providing simple velocity statistics and traffic control rules to the stochastic traffic models, practical vehicular density profile can be readily computed as a function of space and time, which serves as an important clue for predicting various network performance indexes (e.g., BPI, delay and broadcasting throughput) under different signalized traffic scenarios with the developed 802.11p protocol and performance models. The accuracy of the analytical models is validated against extensive simulation, and potential applications of the models to the design of communication protocols and networking functions with enhanced broadcasting performance have been discussed.

Overall, we have presented the viability of combining the traffic models with protocol models to study signalized VANET performance analytically. Further efforts should be put on extending the current set of models to incorporate more complicated road topologies (e.g., 2-D road networks with curvatures and gradients via superposing multiple signalized road segments), probabilistic connectivity and interference models (e.g., with the consideration of fading and attenuation at road-side buildings and street corners), as well as multiple access categories as specified in EDCA for evaluating the delivery of non-safety-related data (such as multimedia data for various infotainment services).

## A. APPENDIX: K-S Test Results for the Homogeneous Traffic Scenario

This appendix contains the K-S test results of comparing the analytical and simulated performance indexes for the homogeneous traffic scenario.

Table A.1. Relative error between the simulated and analytical BPI and the corresponding K-S test results for homogeneous traffic.

| **Lambda** | ***Wss* = 4** | ***Wss* = 8** | ***Wss* = 16** | ***Wss* = 32** | **K-S Test** | ***p* value** |
|---|---|---|---|---|---|---|
| **5** | 0.03 | 0.03 | 0.02 | 0 | 0 | 0.99 |
| **10** | 0.06 | 0.07 | 0.06 | 0.11 | 0 | 0.99 |
| **20** | 0.08 | 0.09 | 0.01 | 0.06 | 0 | 0.99 |
| **30** | 0.06 | 0.07 | 0.06 | 0.05 | 0 | 0.99 |
| **K-S Test** | 0 | 0 | 0 | 0 | | |
| ***p* value** | 0.99 | 0.99 | 0.99 | 0.99 | | |

Table A.2. Relative error between the simulated and analytical delay and the corresponding K-S test results for homogeneous traffic.

| **Lambda** | ***Wss* = 4** | ***Wss* = 8** | ***Wss* = 16** | ***Wss* = 32** | ***Wss* = 64** | **K-S Test** | ***p* value** |
|---|---|---|---|---|---|---|---|
| **2** | 0.08 | 0.02 | 0.03 | 0.06 | 0.08 | 0 | 0.99 |
| **5** | 0.13 | 0.07 | 0 | 0.05 | 0.08 | 0 | 0.99 |
| **10** | 0.13 | 0.02 | 0 | 0.04 | 0.02 | 0 | 0.99 |
| **20** | 0 | 0.03 | 0.01 | 0.02 | 0.04 | 0 | 0.99 |
| **30** | 0.04 | 0.05 | 0.07 | 0.05 | 0.02 | 0 | 0.99 |
| **40** | 0.03 | 0.02 | 0.07 | 0.09 | 0 | 0 | 0.99 |
| **K-S Test** | 0 | 0 | 0 | 0 | 0 | | |
| ***p* value** | 0.81 | 0.81 | 0.99 | 0.99 | 0.99 | | |

Table A.3. Relative error between the simulated and analytical delay and the corresponding K-S test results for homogeneous traffic.

| **Lambda** | ***Wss* = 4** | ***Wss* = 8** | ***Wss* = 16** | ***Wss* = 32** | **K-S Test** | ***p* value** |
|---|---|---|---|---|---|---|
| **2** | 0.07 | 0.08 | 0.26 | 0.43 | 0 | 0.53 |
| **5** | 0.14 | 0.04 | 0.04 | 0.07 | 0 | 0.53 |
| **10** | 0.02 | 0.05 | 0.08 | 0.11 | 0 | 0.99 |
| **20** | 0.06 | 0.06 | 0.08 | 0.01 | 0 | 0.99 |
| **30** | 0.10 | 0.01 | 0.01 | 0.10 | 0 | 0.99 |
| **K-S Test** | 0 | 0 | 0 | 0 | | |
| ***p* value** | 0.99 | 0.99 | 0.99 | 0.70 | | |

## B. APPENDIX: Throughput Estimation for a Fully-Coordinated Broadcast Scheme

In this appendix, an estimate for a fully-coordinated broadcast scheme is presented.

Here, we assume a fully-coordinated scheme that 100% BPI can be achieved at the cost of some extra coordination overhead. The broadcasting throughput with such approach can be preliminarily estimated through (18) – (20) by assuming a coordination overhead of $h$ over a 1 km reception region of average density $n_a$. It is assumed that all cars within the sensing range are coordinated by a central control node. Hence, every car has equal chance to be assigned to any position in the channel access queue, and begins to transmit after all the cars ahead in the queue have finished their transmissions.

Figure B.1 plots the estimated throughput and delay for such scenario assuming zero collision and 20% coordination overhead. The throughput stays at 250 received packets per second, which outperforms the best result achieved by the CSMA/CA case when $W_{ss}$ = 32.

$$t_{frame} = \frac{Payload*(1+h)}{Data\ Rate*16\mu s/slot} \quad (18)$$

$$Delay = t_{frame} * E[\#\ preceding\ trans.] = t_{frame} * \Sigma_{k=0}^{n_a*1} P(k) * k = t_{frame} * \frac{(n_a+1)n_a}{2} * \frac{1}{n_a} \quad (19)$$

$$Throughput = \frac{BPI*n_a*Rs}{Delay*16\mu s/slot} \quad (20)$$

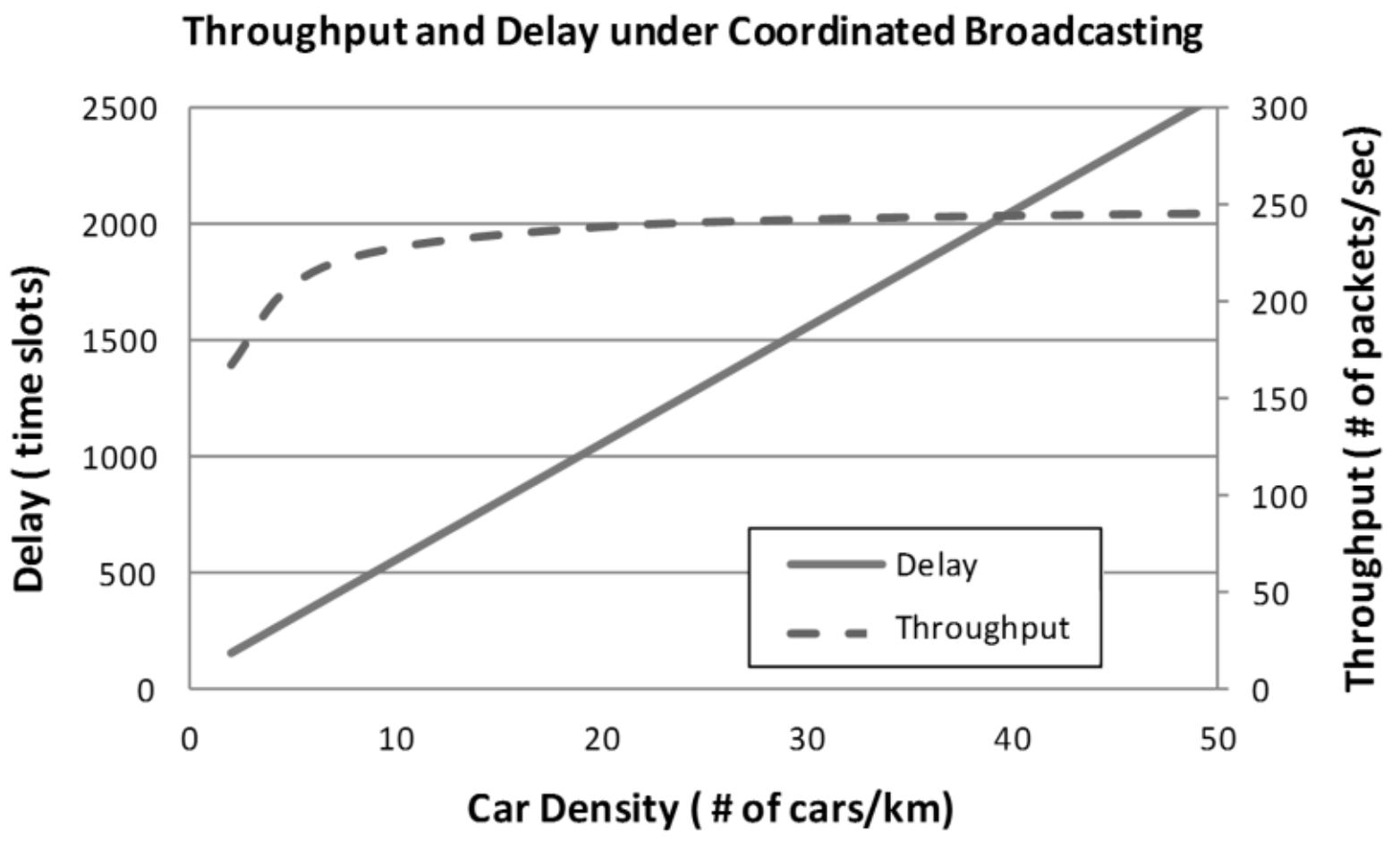

Figure B.1. Throughput and delay with coordinated beaconing (Assuming BPI = 1).

## C. APPENDIX: The Effect of Immediate Access

The channel access of 802.11p is based on CSMA/CA, in which a packet needs to go through a contention process before transmission. One exception is immediate access, where a packet can be transmitted immediately upon arrival [19, 32]. Immediate access happens if the following conditions are met: 1) the channel has been idle for a period longer than AIFS when the new packet arrives; and 2) the message queue is originally empty. In unsaturated traffic scenario, where the communication channel remains mostly idle, both conditions are likely to be fulfilled. Consequently, immediate access could bring non-negligible impact on network performance in the CCH.

If messages other than beacon (e.g., emergency packets) arrive during the CCH interval, which are usually sparse, the vehicular network is likely to operate in an unsaturated state, characterized by more frequent immediate access. Due to the quiet channel, the BPI of the system will remain close to one, while the packet delay drops noticeably after eliminating the contention process. As a result, the throughput is expected to rise by considering immediate access.

To model the immediate access mechanism, the chain model presented in the main text (Figure 3.4) can be modified by splitting the idle state into the immediate access state (IAS) and the non-immediate access state (NIAS) as shown in Figure C.1.

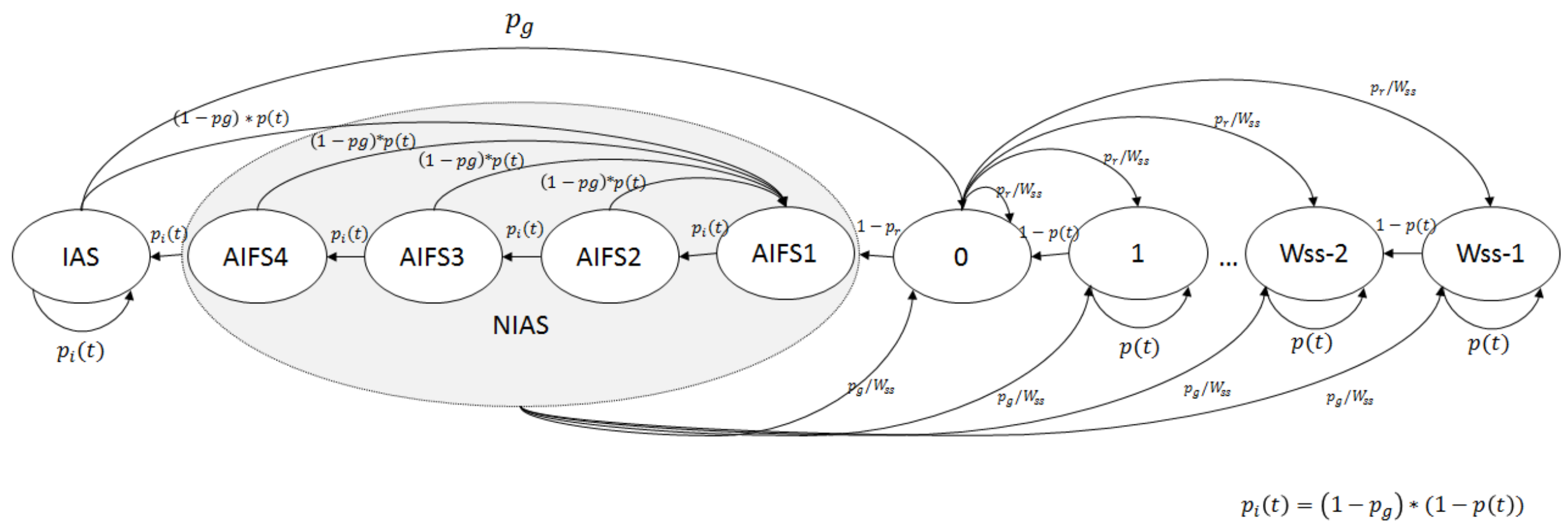

Figure C.1. Modified chain model for describing immediate access under unsaturated condition.

After the broadcasting at state 0, cars enter the non-immediate access state (NIAS), the sub-states in which count the number of consecutive idle timeslots. Each subsequent idle slot sensed advances the system from AIFSn to AIFSn+1 (n = 1, 2, 3). On the other hand, whenever a busy timeslot is sensed, the counter resets to substate AIFS1. Additionally, if an emergency packet arrives (with probability $p_g$) within the NIAS, the node returns to the normal contention process for channel access.

Once a certain number of idle timeslots have been sensed consecutively (i.e., four in our example above), the node switches into the immediate access state (IAS), where a new packet arrival leads to immediate transmission. However, if there is no packet arrival and the channel is sensed busy again, the node goes back to the NIAS as illustrated in Figure C.1. The state transition equations can also be derived from the figure.

As the main effect of immediate access is eliminating the delay overhead brought upon by CSMA/CA contention, we compare the delays predicted by both models (Figure 3.4 and Figure C.1). The scenario chosen is the stable period after the burst of beacon transmission, and emergency packets arrive according to a Poisson arrival process at frequency 10 Hz. In addition, we assume that no packet arrives before the current one is sent out (i.e., $p_r = 0$).

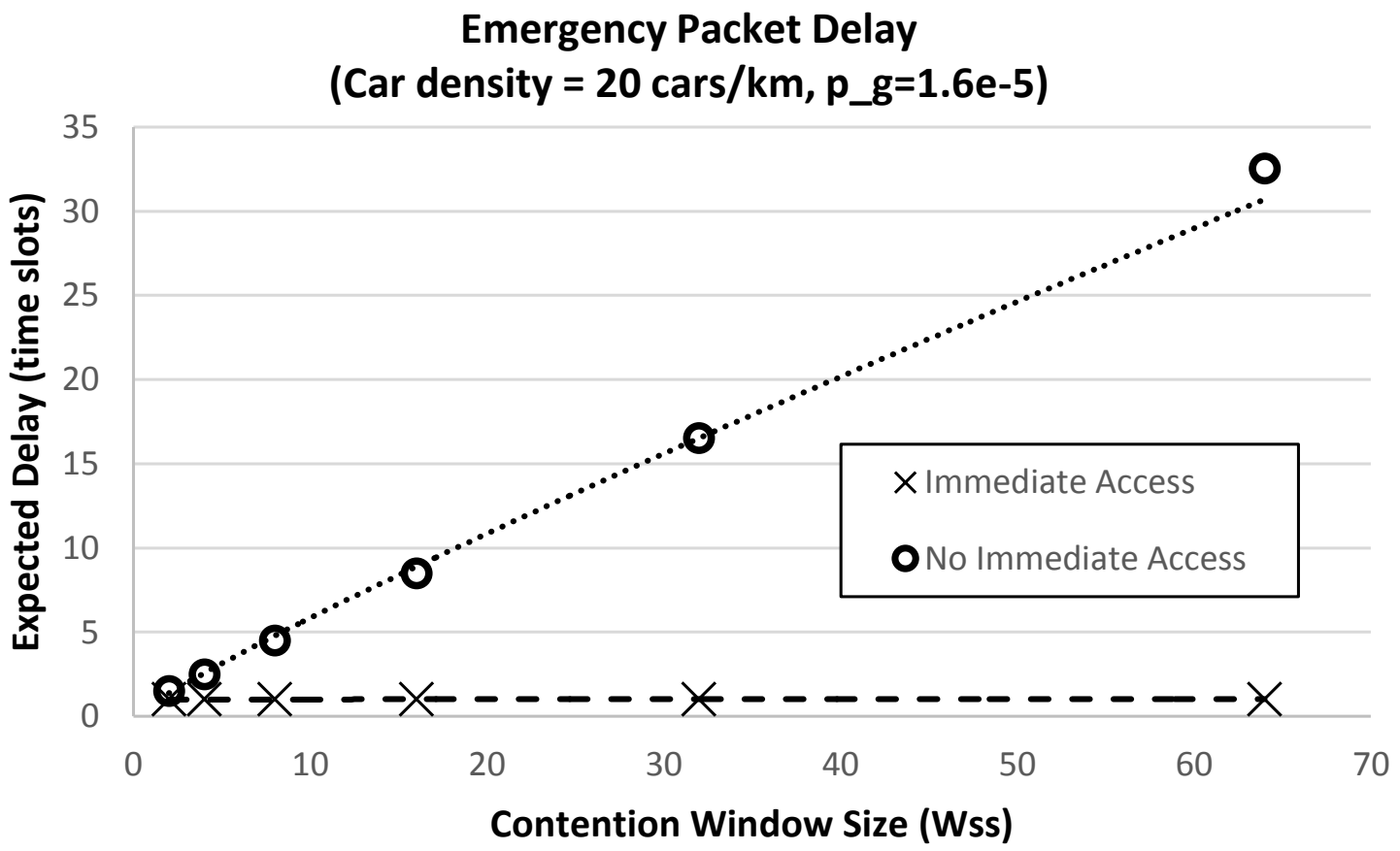

Figure C.2. Analytical comparison of emergency packet delay with and without immediate access.

From Figure C.2, we can see that immediate access keeps delay per packet close to one timeslot; while the average delay without immediate access is around half of the contention window size. As a result, we expect that the throughput would be enhanced noticeably for large *Wss*.

In this example, we have demonstrated that the chain model in Section 3.2.2 can be extended to incorporate immediate access during non-saturated condition. For a more dedicated treatment on immediate access, please refer to [40]. Furthermore, other details can also be added to the basic model presented in the main text to describe the behavior of 802.11p more realistically.

## D. APPENDIX: The Effect of Channel Capturing and Hidden Terminals

Capture effect happens when a node can successfully decode a message amid multiple simultaneously received transmissions, due to a sufficiently high SINR. In this case, we say that the sender of the message successfully captures the channel. As a result, a car falling into the interference range of multiple senders has a chance to receive the right packet given that the desired sender captures the channel.

The main text opted for a simple linear model to describe transmission collision similar to [25, 35], in order to better focus on the relationship between traffic modeling and network performance. However, this approach provides a more conservative throughput and BPI estimation, compared with the case that the capture effect is considered.

Nevertheless, our model can be extended to incorporate the channel capture effect by simply modifying the calculation of $\xi_{a|b,c}$ in (10). In the following, we explain one possible extension as an example.

**1.** Define the relationship between signal propagation and interference. For example, assume that an interferer C corrupts the transmission from sender A to receiver B if the following is satisfied,

$$KP_{tc}d_{AB}^{\alpha} > P_{ta}d_{CB}^{\alpha}. \quad (21)$$

In (21), $K$ is the signal-to-interference requirement, typically 10 dB in ns-2 [41]. $P_{tx}$ is the transmission power of node $x$. If all nodes use the same transmission power, this term can be eliminated. $d_{xy}$ is the Euclidean distance between nodes $x$ and $y$, and $\alpha$ > 2 is the path-loss exponent. If (21) is satisfied, then C will interfere with the transmission from A to B.

Note that here we only introduce how the capture effect could be characterized with a simple free-space propagation model and consider interference from a pair-wise manner. In the future, this could be extended to other non-deterministic signal propagation models and physical interference models that consider aggregated interference from all the simultaneous transmitters in the network.

**2.** Determine the new interference-free target region for the beacon broadcasting based on the signal propagation model. We consider two cases:

**a.** If the effective interferer C is located on the right of (ahead of) the sender A (Figure D.1).

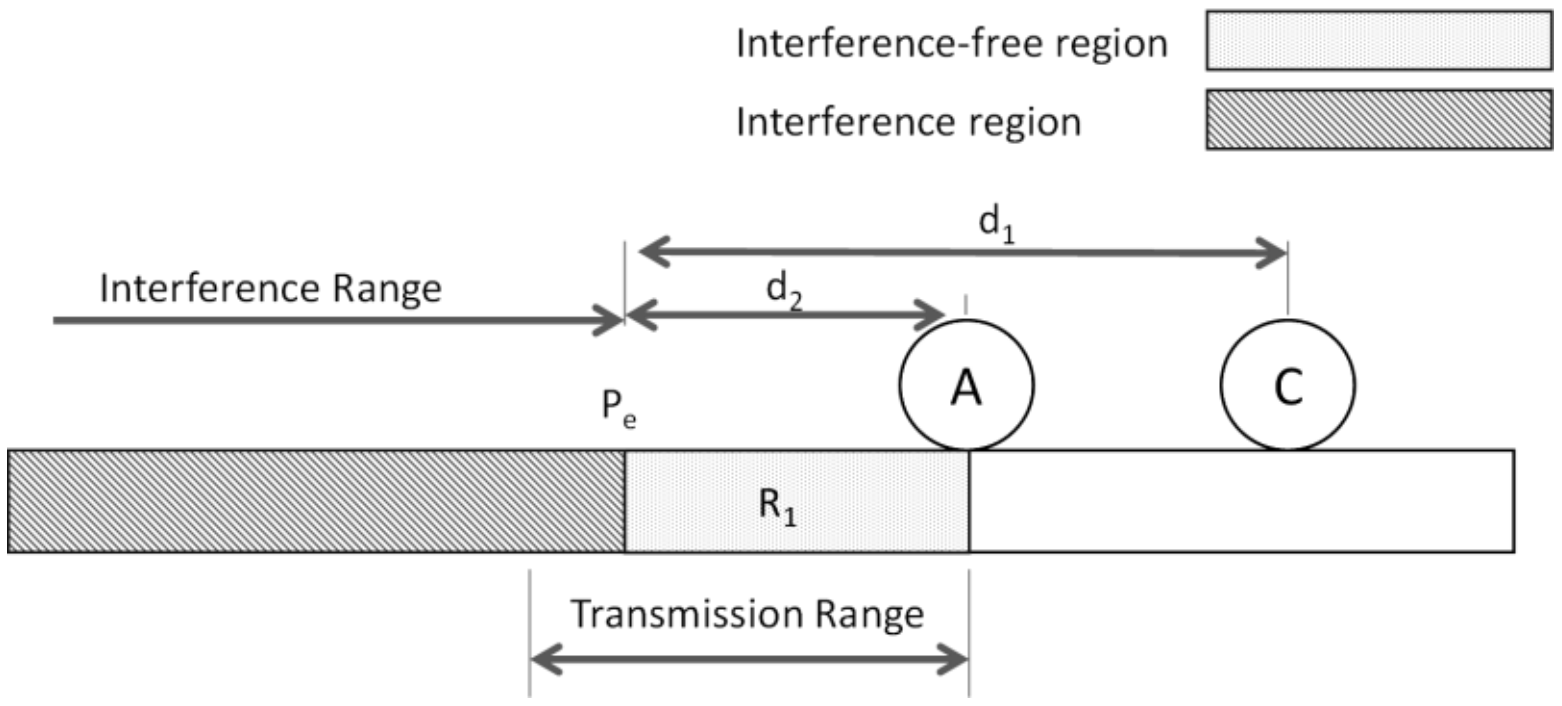

Figure D.1. Illustration of interferer located ahead of the sender (origin on the left, positive direction on the right).

In Figure D.1, there is a point $P_e$ where

$$\boldsymbol{Kd_2^{\alpha} = d_1^{\alpha}}. \quad (22)$$

Locations to the left of $P_e$ will be interfered by C (the interference region) as

$$\boldsymbol{K(d_2 + \Delta d)^{\alpha} > (d_1 + \Delta d)^{\alpha}}. \quad (23)$$

Similarly, locations between $P_e$ and node A are interference-free due to channel captured by A, because we have

$$K(d_2 - \Delta d)^{\alpha} < (d_1 - \Delta d)^{\alpha}. \quad (24)$$

As a result, the interference-free region $R_1$ appears as shown in the figure.

**b.** If the effective interferer is located on the left of (behind) the sender (Figure D.2).

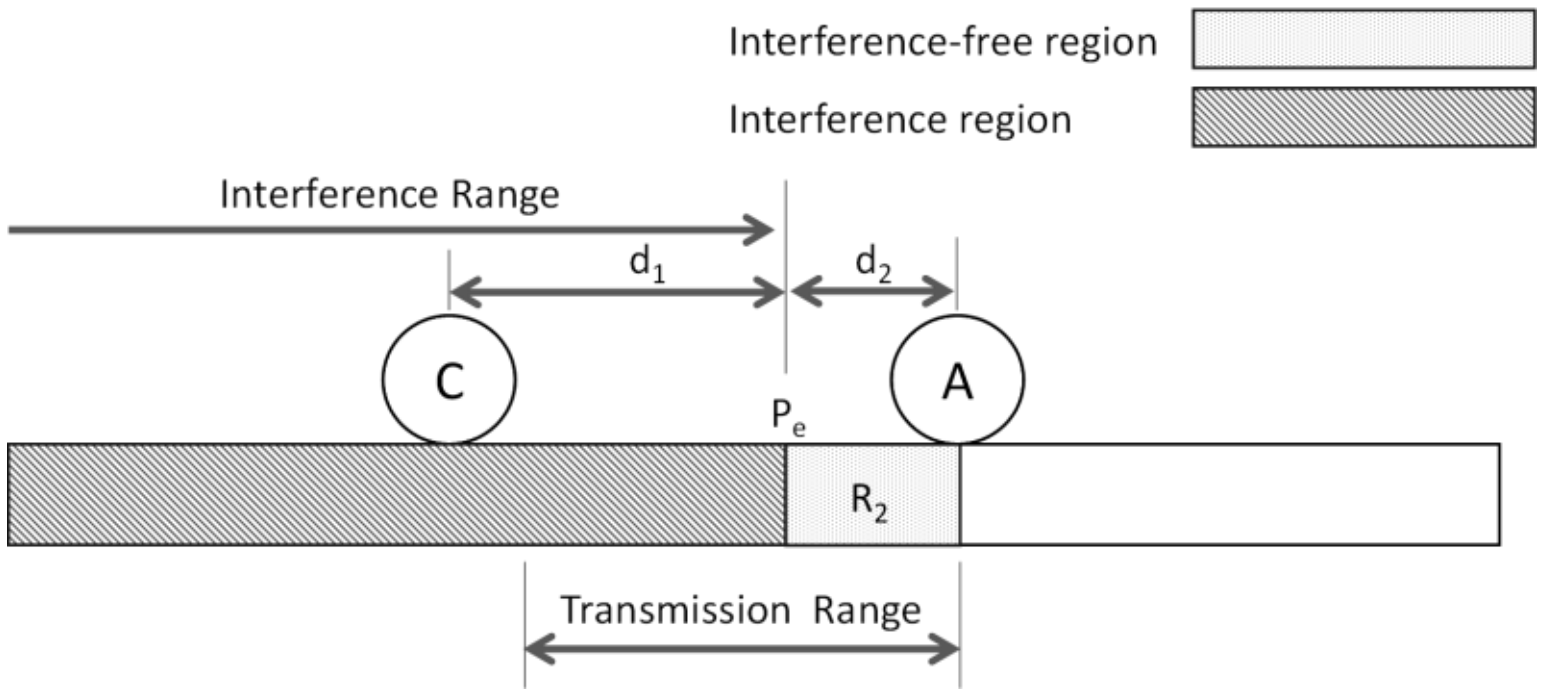

Figure D.2. Illustration of interferer located behind the sender (origin on the left, positive direction on the right).

In Figure D.2, again, there is a point $P_e$ where

$$\boldsymbol{K d_2^{\alpha} = d_1^{\alpha}}. \tag{25}$$

Locations to the left of $P_e$ will be interfered by C (the interference region) as

$$\boldsymbol{K(d_2 + \Delta d)^{\alpha} > (d_1 - \Delta d)^{\alpha}}. \tag{26}$$

Similarly, locations between $P_e$ and node A are interference free due to channel captured by A

$$\boldsymbol{K(d_2 - \Delta d)^{\alpha} < (d_1 + \Delta d)^{\alpha}}. \tag{27}$$

As a result, given the positions of the transmitter and effective interferers, there exists an interference-free region as the intersection of Region 1 ($R_1$) and Region 2 ($R_2$) in the two cases above. Hence, the interference-free target region $R = R_S \cap R_1 \cap R_2$.

**3.** With the newly found interference-free target region, we can re-calculate the average number of cars within the interfered region $\overline{N}_{IR}$ and substitute it into (10) for a more accurate $\xi_{a|b,c}$. When computing $\xi_a$ based on (11), we should integrate from $-\infty$ to *a* for the potential location of the effective interferer *c*.

The above shows that our model can be easily extended to incorporate the capture effect. This approach, however, is less straight forward than the linear interference approach proposed in the main text, and the difference in performance between the two approaches might not be significant when the deviation in the interference-free target region or $\overline{N}_{IR}$ is small. Interested readers can also adopt other signal propagation models according to their specific requirements by calculating and replacing the corresponding $\overline{N}_{IR}$ when determining the broadcasting efficiency $\xi_{a|b,c}$.